\documentclass[letterpaper,english,reprint, aps]{revtex4-2}
\usepackage[T1]{fontenc}
\usepackage[utf8]{inputenc}
\usepackage{babel}
\usepackage{verbatim}
\usepackage{float}
\usepackage{amsmath}
\usepackage{amssymb}
\usepackage{graphicx}
\usepackage{booktabs}
\usepackage{graphicx}
\usepackage{gensymb}
\usepackage{subscript}
\usepackage[unicode=true,pdfusetitle,
bookmarks=true,bookmarksnumbered=false,bookmarksopen=false,
breaklinks=false,pdfborder={0 0 1},backref=false,colorlinks=false]
{hyperref}
\usepackage{xcolor}

\makeatletter

\newcommand{\lyxmathsym}[1]{\ifmmode\begingroup\def\b@ld{bold}
	\text{\ifx\math@version\b@ld\bfseries\fi#1}\endgroup\else#1\fi}

\usepackage{geometry} 
\usepackage{amsmath}
\usepackage{xparse}

\newmuskip\pFqmuskip
\newcommand*\pFq[6][8]{%
	\begingroup 
	\pFqmuskip=#1mu\relax
	\mathchardef\normalcomma=\mathcode`,
	\mathcode`\,=\string"8000
	\begingroup\lccode`\~=`\,
	\lowercase{\endgroup\let~}\pFqcomma
	{}_{#2}F_{#3}{\left[\genfrac..{0pt}{}{#4}{#5};#6\right]}%
	\endgroup
}
\newcommand{\pFqcomma}{{\normalcomma}\mskip\pFqmuskip}

\ExplSyntaxOn
\NewDocumentCommand{\MeijerG}{smmmm}
{
	\IfBooleanTF{#1}
	{
		\vic_meijerg:nnnnnn { #2 } { #3 } { #4 } { #5 } { small } { }
	}
	{
		\vic_meijerg:nnnnnn { #2 } { #3 } { #4 } { #5 } { } { \; }
	}
}

\seq_new:N \l__vic_meijerg_args_in_seq
\seq_new:N \l__vic_meijerg_args_out_seq

\cs_new_protected:Nn \vic_meijerg:nnnnnn
{
	\seq_set_split:Nnn \l__vic_meijerg_args_in_seq { | } { #3 }
	\seq_clear:N \l__vic_meijerg_args_out_seq  
	\seq_map_inline:Nn \l__vic_meijerg_args_in_seq
	{
		\seq_put_right:Nn \l__vic_meijerg_args_out_seq
		{
			\begin{#5matrix} ##1 \end{#5matrix}
		}
	}
	G\sp{#1}\sb{#2}
	\left(
	\seq_use:Nn \l__vic_meijerg_args_out_seq { #6\middle|#6 }
	#6\middle|#6
	#4
	\right)
}
\ExplSyntaxOff

\makeatother
\begin{document}
\title{Anisotropic linear and non--linear excitonic optical properties of buckled monolayer semiconductors}
\author{M. F. C. Martins Quintela$^{1,2,3}$}
\email{mfcmquintela@gmail.com}
\author{T. Garm Pedersen$^{3}$}
\address{$^{1}$Department of Physics and Physics Center of Minho and Porto Universities (CF--UM--UP), Campus of Gualtar, 4710-057, Braga, Portugal}
\address{$^{2}$International Iberian Nanotechnology Laboratory (INL), Av. Mestre
	Jos{\'e} Veiga, 4715-330, Braga, Portugal}
\affiliation{$^{3}$Department of Materials and Production, Aalborg University, 9220 Aalborg {\O}st, Denmark}
\begin{abstract}
	The optical properties of two-dimensional materials are exceptional in several respects. They are highly anisotropic and frequently dominated by excitonic effects. Dipole--allowed second order non--linear optical properties require broken inversion symmetry. Hence, several two--dimensional materials show strong in--plane (IP) non--linearity but negligible out--of--plane (OOP) response due to vertical symmetry. By considering buckled hexagonal monolayers, we analyze the critical role of broken vertical symmetry on their excitonic optical response. Both linear as well as second order shift current and second harmonic response are studied. We demonstrate that substantial OOP non--linear response can be obtained, in particular, through off--diagonal tensor elements coupling IP excitation to OOP response. Our findings are explained by excitonic selection rules for OOP response and the impact of dielectric screening on excitons is elucidated.
\end{abstract}
\maketitle

\section{Introduction}

The recent interest in layered materials with broken vertical symmetry, such as Janus materials\cite{Lu2017,Zhang2017,doi:10.1021/acs.nanolett.0c03412}, buckled monolayers\cite{siahin_monolayer_2009,PhysRevB.99.085432,le_fracture_2021,kezerashvili_effects_2021}, as well as heterobilayers and biased homobilayers\cite{doi:10.1021/nl902932k,Gong2014,Rivera2016,PhysRevB.91.205405} makes the discussion on the effects of broken vertical symmetry on the optical response especially relevant\cite{doi:10.1021/acs.nanolett.7b05286,Niu2018,Brotons-Gisbert2019,PhysRevLett.127.076402,Ermolaev2021,PhysRevB.106.054408,PhysRevB.106.075143}. The amplitude of both linear and non--linear OOP conductivities is expected to be greatly dependent on the asymmetry of the layer, with the even--order non--linear OOP response being identically zero (within the dipole approximation) when the OOP symmetry is not broken. Hence, the broken OOP symmetry is crucial when one wishes to consider potential applications beyond those allowed by symmetric structures.
The OOP non--linear response in Janus monolayers has also been experimentally studied\cite{arxiv.2303.03844,shi2023giant}, namely for both second-- and third--harmonic. This study was performed via polarization--resolved spectroscopy, with the aim of mapping the full second--order susceptibility tensor\cite{C9CP03395E,doi:10.1021/acs.nanolett.2c00898,doi:10.1021/acs.jpcc.2c03792} of $\mathrm{MoSSe}$. These OOP non--linearities then lead to additional degrees of freedom in vertical photonics structures\cite{Kleemann2017,doi:10.1021/acs.nanolett.8b02652}, allowing for novel approaches in the design of ultrafast optical devices\cite{https://doi.org/10.1002/lpor.202100726}, such as miniaturized logic gates\cite{doi:10.1126/sciadv.abq8246,Li2022}, non--linear holograms\cite{doi:10.1021/acs.nanolett.9b02740}, broadband ultrafast frequency converters\cite{Soavi2018,Klimmer2021}, among others. 

\begin{figure}
	\centering{}\includegraphics[scale=0.3]{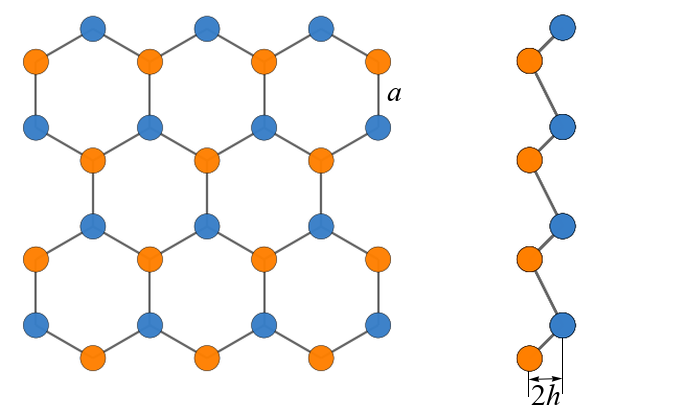}
	\caption{Schematic illustration of a buckled honeycomb lattice, highlighting the lattice constant $a$ and the buckling $h$.  \label{fig:lattice}}
\end{figure}

The simplest family of materials with broken OOP symmetry is that of buckled monolayer structures, with theoretical predictions of both mono--elemental and binary graphene--like materials\cite{siahin_monolayer_2009,le_fracture_2021,kezerashvili_effects_2021}, and several buckled hexagonal sheets (see Fig. \ref{fig:lattice}) have already been fabricated. Among these materials we mention specifically the mono--elemental silicene\cite{siahin_monolayer_2009,takeda_theoretical_1994}, blue phosphorene\cite{zhu_semiconducting_2014}, arsenene\cite{zhang_atomically_2015,kamal_arsenene_2015}, antimonene\cite{zhang_atomically_2015,wang_atomically_2015}, and bismuthene\cite{zhang_semiconducting_2016}, as well as the binary $ \mathrm{CS} $, $ \mathrm{SiO}$, $ \mathrm{GeSe} $, $ \mathrm{SnTe} $, $ \mathrm{InSb} $, and $ \mathrm{GaAs} $\cite{siahin_monolayer_2009,kamal_direct_2016}. The mono--elemental structures preserve inversion symmetry even in the presence of buckling and, hence, possess negligible second--order non--linearities. The bandgaps of these materials can be both mechanically\cite{yan_tuning_2015,molle_buckled_2017} or electrically\cite{molle_buckled_2017,ni_tunable_2012} tuned, and they allow for potential applications is various fields, such as optoelectronics, 
spin-electronics,
sensors and thermo-electrics\cite{vishnoi_2d_2019,geng_recent_2018,wang_2d_2018,liu_recent_2019}.

The aim of the present work is to understand the effects of both IP and OOP asymmetry on the excitonic optical response of honeycomb lattice structures. To this end, we consider a simple two--band model of gapped graphene near the so--called Dirac valleys\cite{novoselov_two-dimensional_2005,castro_neto_electronic_2009,stauber_optical_2008} and then apply a small buckling to break OOP symmetry. To study IP even--order non--linear optical properties\cite{ventura_gauge_2017,passos_nonlinear_2018,boyd_nonlinear_2020}, such as second--harmonic generation (SHG)\cite{sie_coherent_2018,boyd_nonlinear_2020} or shift--current (SC)\cite{moore_confinement-induced_2010,young_first_2012,sodemann_quantum_2015}, we include a quadratic (in $k$) contribution to the nearest neighbour hopping function\cite{ventura2021nonlinear}, namely trigonal warping\cite{taghizadeh_nonlinear_2019}, plus distinct on--site potentials for the two sublattices. Including trigonal warping allows us to compute the IP even--order response, which then serves as a comparison against the OOP response.

This paper is organized as follows. In Section \ref{Sec:gapped}, we will consider the single--particle Hamiltonian for gapped graphene while introducing trigonal warping before computing explicit matrix elements of the momentum and Berry connection. In Section \ref{Sec:BSE}, we discuss the Bethe--Salpeter equation for the computation of the excitonic states. We also outline some of the approximations necessary for an efficient numerical solution of this equation. In Section \ref{Sec:optical_response_1}, we briefly outline the general form of both the excitonic linear and non--linear optical response to linearly polarized light, discussing the momentum matrix elements between excitonic states. Finally, in Section \ref{Sec:buckling}, we analyze the IP and OOP optical selection rules of a buckled graphene lattice structure (Fig. \ref{fig:lattice}) which, in turn, leads to a non--zero OOP excitonic response in the monolayer. The non--linear response will be very sensitive to the scale of this buckling, quickly vanishing as the buckling decreases. We also compare both diagonal $\left(\sigma_{zzz}\right)$ and non--diagonal  $\left(\sigma_{zxx/xzx}\right)$ components of the second order excitonic conductivity tensor against their IP counterparts, discussing both their relative magnitudes and the location of the excitonic resonances. 

\section{Single Particle Gapped Graphene Hamiltonian}\label{Sec:gapped}

Throughout this paper, we will work with a two--band model of gapped graphene near the Dirac points $K/K^\prime$, with the $x$--axis aligned with the unit cell of the honeycomb lattice and the $z$--axis perpendicular to the monolayer plane. 
The basis states are $p_z$ orbitals on the two sublattices and the model Hamiltonian before Dirac point expansion then reads
\begin{equation}
\mathcal{H}\left(\mathbf{k}\right)=\left[\begin{array}{cc}
\Delta & -\gamma f^{*}\left(\mathbf{k}\right)\\
-\gamma f\left(\mathbf{k}\right) & -\Delta
\end{array}\right],\label{eq:Ham}
\end{equation}
where $\pm\Delta$ is the staggered on--site energy and $\gamma$ the effective hopping. While for planar gapped graphene the $\pi$ and $\sigma$ orbitals are decoupled, the vertical shift of the two sublattices in a buckled system means that the $p_z$ orbitals are no longer in the same plane. Hence, the effective hopping will change as\cite{PhysRev.94.1498,Song_2018} 
\begin{align}
	-\gamma=V_{p p \pi}+\frac{1}{1+\frac{a^2}{12 h^2}}\left(V_{p p \sigma}-V_{p p \pi}\right),\label{eq:hybridization}
\end{align}
where $ a $ is the lattice parameter, $h$ is the buckling parameter, and $V_{p p \pi}$ and $V_{p p \sigma}$ are the hopping integrals for $\pi$ and $\sigma$ orbitals, respectively. 
Additionally, we will ignore $\pi-\sigma$ hybridization when computing the OOP response as we consider the OOP buckling to be much smaller than the lattice constant. 
The wave--vector dependent function $f$ is obtained from the
honeycomb lattice geometry as 
\[
f\left(\mathbf{k}\right)=e^{i\frac{k_{x}a}{\sqrt{3}}}+2e^{-i\frac{k_{x}a}{2\sqrt{3}}}\cos\left(\frac{k_{y}a}{2}\right).
\]

Expanding $f\left(\mathbf{k}\right)$ near the Dirac points $K/K^{\prime}$
up to linear order, we obtain the massive Dirac Hamiltonian that is
usually employed to study gapped graphene and hBN systems. Considering
now an expansion up to quadratic order in $k$, we obtain \cite{taghizadeh_nonlinear_2019}
\begin{equation}
f\left(\mathbf{k}\right)\approx\frac{\sqrt{3}a}{2}\left[\left(k_{x}+i\tau k_{y}\right)+i\zeta_{\mathrm{TW}}a\left(k_{x}-i\tau k_{y}\right)^{2}\right],\label{eq:trig_warp_f}
\end{equation}
where $\tau=\pm1$ is the valley index and $\zeta_{\mathrm{TW}}=\frac{\sqrt{3}}{12}$ is the trigonal warping strength. Although this trigonal warping strength is a fixed numerical factor, it is useful to keep it as a variable to enable systematic expansions in orders of $\zeta_{\mathrm{TW}}$.

\subsection{Diagonalization}

Diagonalizing the Hamiltonian Eq. (\ref{eq:Ham}), we obtain the
band structure as $	\pm E$ with
	\begin{align}
	E&=\sqrt{\Delta^{2}+\gamma^2\left|f\left(\mathbf{k}\right)\right|^2}.
\end{align}
As we are interested in linear contributions from trigonal warping, we approximate $E$ up to first order in $ \zeta_{\mathrm{TW}} $ as 
\begin{equation}
	E\approx\varepsilon+\tau\frac{\xi}{\varepsilon}\zeta_{\mathrm{TW}},\label{eq:bandstructure-fixed}
\end{equation}
where
\begin{align}
	\varepsilon & =\sqrt{\Delta^{2}+\hbar^{2}v_F^{2}k^{2}},\nonumber\\
	\xi & =a \hbar^{2}v_F^{2} k^{3}\sin\left(3\theta\right),\label{eq:short_var}
\end{align}
and the Fermi velocity is defined $v_{F}=\frac{1}{\hbar}\frac{\sqrt{3}a\gamma}{2}$.
We then write the normalized eigenvectors as
\begin{align}
		\left|v_{\mathbf{k}}\right\rangle  & =\sqrt{\frac{E+\Delta}{2E}}\left[\begin{array}{c}
		\frac{e^{-i\tau\theta}\left(E-\Delta\right)}{\hbar v_F k\left(1 + i a k \zeta_{\mathrm{TW}} e^{-3 i \theta \tau}\right)}\\
		1
		\end{array}\right],\label{eq:eigenvectors_fixed1}\\
	\left|c_{\mathbf{k}}\right\rangle  & =\sqrt{\frac{E-\Delta}{2E}}\left[\begin{array}{c}
	\frac{-E-\Delta}{\hbar v_F k\left(1 + i a k \zeta_{\mathrm{TW}} e^{-3 i \theta \tau}\right)}\\
		e^{i\tau\theta}
	\end{array}\right],\label{eq:eigenvectors_fixed2}
\end{align}
where $v/c$ correspond to the valence and conduction band, respectively. From Eqs. (\ref{eq:eigenvectors_fixed1}-\ref{eq:eigenvectors_fixed2}) it is clear which components go to zero as $k\rightarrow0$, as $E\approx \Delta + \mathcal{O}\left(k^2\right)$ for small $k$, whereas the denominators of the fraction in square brackets are $\mathcal{O}(k)$. 

The presence of the phase terms in spinor components that go to zero as $k\rightarrow0$ in $ \left|v_{\mathbf{k}}\right\rangle$ and $\,\left|c_{\mathbf{k}}\right\rangle $ will lead to a pseudo-spin angular quantum number $ m_s=0 $\cite{park_tunable_2010,henriques_absorption_2022,di_sabatino_optical_2020,zhang_optical_2018,cao_unifying_2018,henriques_excitonic_2022}. This pseudo-spin angular quantum number is governed by the phase choice and allows a direct association with the usual Hydrogen--like states. 

\subsection{Momentum Matrix Element and Berry Connection}

The IP interband momentum matrix element in the $i$--direction is defined as
\begin{align}
p_{vc\mathbf{k}}^{i} & 
 =\left\langle v_{\mathbf{k}}\middle|\frac{m}{\hbar}\frac{\partial\mathcal{H}\left(\mathbf{k}\right)}{\partial k_{i}}\middle|c_{\mathbf{k}}\right\rangle.\label{eq:momentum_general}
\end{align}
When considering IP properties, we will focus our attention solely on the $ x $--direction as the inversion symmetry of the lattice along the $y$--direction means that the $yyy$--component of the non--linear conductivity tensor will be trivially zero after summing over valley index. 
Alongside with the momentum matrix elements, we will also require Berry connections, defined as 
\begin{equation}
	\Omega_{nm\mathbf{k}}^{\alpha}=i\left\langle n_\mathbf{k}\left|\frac{\partial}{\partial k_{\alpha}}\right|m_\mathbf{k}\right\rangle \label{eq:berry_connection-before}
\end{equation}
as their explicit expression will play an important part in computing generalized derivatives\cite{aversa_nonlinear_1995}. 

To obtain the non--linear conductivity tensor, we will consider incident fields with frequency $\omega_p$ and $\omega_q$.
The indices for the current vector $\mathbf{J}^{(2)}\left(\omega_{p q}\right)$ will contract as\cite{boyd_nonlinear_2020}
\begin{equation}
	J_i^{(2)}(\omega_{p q})=\sum_{j, k} \sigma_{i j k}^{(2)}(\omega_{p q} ; \omega_p, \omega_q) E_j(\omega_p) E_k(\omega_q)
\end{equation}
with $\mathbf{E}\left(\omega\right)$ the external optical field and the frequency $\omega_{p q}=\omega_p + \omega_q$. A simple symmetry analysis\cite{arxiv.2303.03844}
tells us that the relevant components of the non--linear conductivity tensor will be
\begin{equation*}
	\sigma^{\left(2\right)}=\left[\begin{array}{cccccc}
		\sigma_{xxx}^{(2)} & -\sigma_{xxx}^{(2)} & 0 & 0 & \sigma_{xxz}^{(2)} & 0\\
		0 & 0 & 0 & \sigma_{xxz}^{(2)} & 0 & -\sigma_{xxx}^{(2)}\\
		\sigma_{zxx}^{(2)} & \sigma_{zxx}^{(2)} & \sigma_{zzz}^{(2)} & 0 & 0 & 0
	\end{array}\right].
\end{equation*}

Compared to \cite{taghizadeh_nonlinear_2019}, we apply a simple relabeling of the two valleys $ \left(\tau\rightarrow-\tau\right) $ and the gauge change $ \left|c_{\mathbf{k}}\right\rangle \rightarrow e^{i\tau\theta}\left|c_{\mathbf{k}}\right\rangle $. While this gauge change leads to a global phase in the momentum matrix elements, it is important to note that the $k_x$--derivative in the definition of the Berry connection will lead to a more complex transformation. Nonetheless, this is just a gauge choice and, therefore, both the free--carrier and the excitonic conductivity will be independent of this choice.

\subsection{Free--Carrier Conductivity}

When discussing both linear and non--linear excitonic conductivities, we will include results for very large dielectric constants. In this limit, the excitonic response agrees with the free--carrier expression, obtained by computing the electronic conductivity in the free--carrier (single particle) regime. 

The generic expression for the free--carrier linear electronic conductivity in a clean two--band semiconductor at $T=0$ is given by \cite{genkin1968contribution,aversa_nonlinear_1995,kirtman_extension_2000,margulis_optical_2013,pedersen_intraband_2015}
\begin{align}
	\sigma_{\alpha \beta}\left(\omega\right)
	 & =\frac{e^2 \hbar}{i \pi^2 m^2} \left[\int \frac{p_{vc\mathbf{k}}^\alpha p_{cv\mathbf{k}}^\beta}{E_{cv\mathbf{k}}\left(E_{cv\mathbf{k}}-\hbar \omega\right)} d^2 \mathbf{k}\right.\nonumber\\
	 &\quad\left.-\left(\omega\rightarrow-\omega\right)^*\right], \label{eq:generic_electronic}
\end{align}
where $ E_{cv\mathbf{k}}=2E $ and the integration runs over the Brillouin zone. Analogously, the generic intraband non--linear electronic conductivity in a clean two--band semiconductor can be written as \cite{genkin1968contribution,aversa_nonlinear_1995,kirtman_extension_2000,margulis_optical_2013,pedersen_intraband_2015}
\begin{widetext}
	\begin{align}
		\sigma_{\alpha \beta \lambda}^{(\mathrm{i n t r a})}\left(\omega_p, \omega_q\right)=\frac{e^3 \hbar^2\left(\omega_p+\omega_{p q}\right)}{2 \pi^2 m^2} \int \frac{p_{v c\mathbf{k}}^\alpha\left[p_{c v\mathbf{k}}^\beta\right]_{; k_\lambda}}{\left(E_{cv\mathbf{k}}^2-\hbar^2 \omega_p^2\right)\left(E_{cv\mathbf{k}}^2-\hbar^2 \omega_{p q}^2\right)} d^2 \mathbf{k}+(p \leftrightarrow q),\label{eq:generic_electronic-nl}
	\end{align}
\end{widetext}
where $ \left[p_{c v\mathbf{k}}^\beta\right]_{; k_\lambda} $ is the generalized derivative\cite{aversa_nonlinear_1995} in the $ \lambda $--direction of the momentum matrix element for the $ \beta $--direction, defined as
\begin{align}
	\left[p_{c v\mathbf{k}}^\beta\right]_{; k_\lambda}&=\frac{\partial p_{c v\mathbf{k}}^\beta}{\partial k_\lambda}-i\left(\Omega_{cc\mathbf{k}}^{\lambda}-\Omega_{vv\mathbf{k}}^{\lambda}\right)p_{c v\mathbf{k}}^\beta.\label{eq:generalized_deriv-momentum}
\end{align}
When considering $\lambda=z$, the $k_z$--derivative term in Eq. (\ref{eq:generalized_deriv-momentum}) is discarded as there is no dependence on $k_z$ in the momentum matrix elements. The specific details for the calculation of both $p_{vc\mathbf{k}}^{z}$ and $\Omega_{nm\mathbf{k}}^{z}$ will be discussed in Section \ref{Sec:buckling}. 

While the integrals of Eqs. (\ref{eq:generic_electronic}-\ref{eq:generic_electronic-nl}) are over the entire Brillouin zone, performing the expansion around the Dirac points means that the integration is now over the infinite Dirac cone and that a sum over valleys must be made. Due to the smallness of $\zeta_{\mathrm{TW}}$, we are interested contributions up to $\mathcal{O}(\zeta_{\mathrm{TW}})$. The $\zeta_{\mathrm{TW}}$ factor must come from either $E_{cv\mathbf{k}}$ or $ p_{v c\mathbf{k}}^\alpha\left[p_{c v\mathbf{k}}^\beta\right]_{; k_\lambda}$. Time reversal symmetry means that $E_{cv\mathbf{k}}=E_{cv-\mathbf{k}}$. Any term containing $p_{v c\mathbf{k}}^\alpha\left[p_{c v\mathbf{k}}^\beta\right]_{; k_\lambda}$ to zeroth order in $\zeta_{\mathrm{TW}}$ will vanish upon integration and summation over valley. This allows us to set $E_{cv\mathbf{k}}=2\varepsilon $ while retaining $\mathcal{O}\left(\zeta_{\mathrm{TW}}\right)$ contribution to $p_{v c\mathbf{k}}^\alpha\left[p_{c v\mathbf{k}}^\beta\right]_{; k_\lambda}$ throughout this paper when computing the various transition amplitudes. These integrals can be computed analytically in our first--order approximation in $ \zeta_{\mathrm{TW}} $, with the exact expressions present in Appendix \ref{app:electronic_conductivities} for the various processes considered.

\section{Bethe--Salpeter Equation}\label{Sec:BSE}

Before discussing the excitonic conductivity, we must first compute the excitonic states for each $\tau$ valley. To compute the excitonic wave functions and their binding energies, we will solve the Bethe--Salpeter equation\cite{pedersen_intraband_2015,taghizadeh_nonlinear_2019,cao_unifying_2018,radha_optical_2021}, given in momentum space by
\begin{equation}
	\begin{aligned}
	&E_n\psi_{cv\mathbf{k}}^{\left(n\right)}=  E_{cv\mathbf{k}}\psi_{cv\mathbf{k}}^{\left(n\right)}+\\
	&	+\sum_{\mathbf{q}}V\left(\left|\mathbf{k}-\mathbf{q}\right|\right)\left\langle c_{\mathbf{k}}\middle| c_{\mathbf{q}}\right\rangle \left\langle v_{\mathbf{q}}\middle|v_{\mathbf{k}}\right\rangle \psi_{cv\mathbf{q}}^{\left(n\right)}\end{aligned}
	,\label{eq:BSE}
\end{equation}
where $E_n$ is the exciton energy of state $n$, $V\left(k\right)$ is the
attractive electrostatic potential coupling electrons and holes,
and $\psi_{cv\mathbf{k}}^{\left(n\right)}$ is the wave function of the
exciton. For notational simplicity, the $\tau$ dependence of energy and wave function is omitted from the list of arguments. In Eq. (\ref{eq:BSE}), the valley dependence is present in the form factor $\left\langle c_{\mathbf{k}}\middle| c_{\mathbf{q}}\right\rangle \left\langle v_{\mathbf{q}}\middle|v_{\mathbf{k}}\right\rangle$. For our system, we consider $V\left(k\right)$ to be the
Rytova--Keldysh potential\cite{rytova_screened_1967,keldysh_coulomb_1979}, given in momentum space by 
\begin{equation}
	V\left(k\right)=-2\pi\hbar c\alpha\frac{1}{k\left(\epsilon+r_{0}k\right)},
\end{equation}
with $\alpha$ the fine--structure constant, $\epsilon$ the mean
dielectric constant of the media surrounding the monolayer, and $r_{0}$
an IP screening length\cite{li_excitons_2019} related to the polarizability of
the material and usually obtained from DFT calculations\cite{tian_electronic_2020}. From the analysis of Fig. 2C of Ref. \cite{tian_electronic_2020} for graphene--like materials with a bandgap of $E_g=2\,\mathrm{eV}$, we set $r_{0}=40\,\text{\AA}$.

Considering the excitonic wave function to have a well--defined
angular momentum $\ell_n$, we write it as $\psi_{cv\mathbf{k}}^{\left(n\right)}=f_{cvk}^{\left(n\right)}e^{i\ell_n\theta_{k}}$ and, defining $\varphi=\theta_{q}-\theta_{k}$, rewrite the Bethe--Salpeter equation by converting the sum into an integral as 
\begin{widetext}
	\begin{align}
		E_n f_{cvk}^{\left(n\right)} & =2\varepsilon f_{cvk}^{\left(n\right)}+\frac{1}{4\pi^{2}}\sum_{\lambda=0}^{2}\int_{0}^{\infty} \int_{0}^{2\pi} \,V\left(\left|\mathbf{k}-\mathbf{q}\right|\right)\mathcal{A}_{\lambda}\left(k,q\right)e^{i\lambda\tau\varphi}f_{cvq}^{\left(n\right)}e^{i\ell_n\varphi}d\varphi\,qdq,\label{eq:bse_full-1}
	\end{align}
\end{widetext}
where $E_{cv\mathbf{k}}$ became $2\varepsilon$ as we are neglecting the effects of trigonal warping on the band structure for simplicity. This approximation removes all coupling of states with different angular momentum. 

The radial component of the form factor is obtained directly from the expansion of $\left\langle c_{\mathbf{k}}\middle| c_{\mathbf{q}}\right\rangle \left\langle v_{\mathbf{q}}\middle|v_{\mathbf{k}}\right\rangle$ while again neglecting trigonal warping in the definition of the eigenvectors. Under this approximation, the eigenvectors read
\begin{align}
	\left|v_{\mathbf{k}}\right\rangle  & =\left[\begin{array}{c}
		e^{-i\tau\theta}\sin\frac{x_{k}}{2}\\
		\cos\frac{x_{k}}{2}
	\end{array}\right],\nonumber&
	\left|c_{\mathbf{k}}\right\rangle  & =\left[\begin{array}{c}
		-\cos\frac{x_{k}}{2}\\
		e^{i\tau\theta}\sin\frac{x_{k}}{2}
	\end{array}\right],\label{eq:eigenvectors_fixed-simp-final}
\end{align}
where $x_{k}=\tan^{-1}\left[\frac{\hbar v_{F} k}{\Delta}\right]$. The radial component of the form factor can then be written as
\[
\mathcal{A}_{\lambda}\left(k,q\right)=\begin{cases}
	\frac{1}{4}\left(1+\cos x_{k}\right)\left(1+\cos x_{q}\right), & \lambda=0\\
	\frac{1}{2}\sin x_{k}\sin x_{q}, & \lambda=1\\
	\frac{1}{4}\left(1-\cos x_{k}\right)\left(1-\cos x_{q}\right), & \lambda=2
\end{cases},
\]
where $\lambda$ denotes the angular dependence present in the $ e^{i\lambda\tau\varphi} $ factor in Eq. (\ref{eq:bse_full-1}). 

As is evident from Eq. (\ref{eq:bse_full-1}), the degeneracy in angular momentum $ \ell_n\leftrightarrow-\ell_n $ is immediately lifted within the same valley. However, a degeneracy between $ \left(\ell_n+m_s,\tau\right) $ and $ \left(-\ell_n-m_s,-\tau\right) $ excitons is still present, stemming from time reversal symmetry in the system\cite{wu_exciton_2015}. Finally, Eq. (\ref{eq:bse_full-1}) is solved numerically via a simple numerical quadrature using a tangent grid $k=\tan\left(x\frac{\pi}{2}\right)$ with $1000$ points $x\in\left[0,1\right]$, following the procedure already outlined several times in the literature, namely in Refs. \cite{chao_analytical_1991,parfitt_two-dimensional_2002,henriques_absorption_2022,quintela_tunable_2022}. 

When discussing excitonic states, we will use nomenclature similar to the 2D Hydrogen atom to distinguish the different angular momentum states (\emph{i.e.}, $ s $, $ p_\pm $, $ d_\pm $ states). As the pseudo--spin contribution $m_s=0$, $s$--states will have $\ell=0$, $p_\pm$ will have $\ell=\pm1$, and analogously to higher angular momentum states.

\section{Excitonic Optical Response}\label{Sec:optical_response_1}

Having outlined the method for obtaining excitonic states in our system, we will now consider the excitonic optical conductivity. To extract the resonances, we add a broadening of $ \hbar\Gamma=0.05\,\mathrm{eV} $, introduced via the substitution $ \hbar\omega\rightarrow \hbar\omega+i\frac{\hbar\Gamma}{2} $. Following \cite{taghizadeh_gauge_2018,taghizadeh_nonlinear_2019,pedersen_intraband_2015,mkrtchian_theory_2019}, we define $ \sigma_{0}=\frac{e^2}{4\hbar} $ and write the linear conductivity  as
\begin{equation}
	\frac{\sigma_{\alpha\beta}\left(\omega\right)}{\sigma_{0}}=\frac{-i\hbar^2}{2\pi^3 m^2} \sum_n\left[\frac{E_n X_{0 n}^\alpha X_{n 0}^\beta}{E_n-\hbar \omega}-\left(\omega\rightarrow-\omega\right)^*\right].\label{eq:excitonic_first-order_simp1}
\end{equation}
For the non--linear conductivity, we are interested in both SC $\left(\omega_p=-\omega_q^*\right)$ and SHG $\left(\omega_p=\omega_q\right)$ regimes. Defining $ \sigma_2=\frac{e^3 a }{4 E_g\hbar} $, we write the SHG non--linear conductivity\cite{pedersen_intraband_2015,taghizadeh_nonlinear_2019} as
\begin{widetext}
	\begin{equation}
		\frac{\sigma_{\alpha\beta\gamma}^{\mathrm{SHG}}\left(\omega\right)}{\sigma_2}=\frac{-iE_g \hbar^2}{2a \pi^3 m^2} \sum_{n, m}\left[\frac{E_n X_{0 n}^\alpha Q_{n m}^\beta X_{m 0}^\gamma}{\left(E_n-2 \hbar \omega\right)\left(E_m-\hbar \omega\right)}-\frac{E_n X_{n 0}^\alpha Q_{m n}^\beta X_{0 m}^\gamma}{\left(E_n+2 \hbar \omega\right)\left(E_m+\hbar \omega\right)}-\frac{\left(E_n-E_m\right) X_{0 n}^\alpha  Q_{n m}^\beta X_{m 0}^\gamma}{\left(E_n+\hbar \omega\right)\left(E_m-\hbar \omega\right)}\right].\label{eq:excitonic_second-order_simp1}
	\end{equation}
\end{widetext}
In these expressions, $ E_n $ is the energy of the excitonic state $ n $, and the one-- and two--state excitonic matrix elements are defined as \cite{pedersen_intraband_2015,taghizadeh_nonlinear_2019}
\begin{equation}
	X_{0 n}^\alpha =i \int \,\psi_{cv\mathbf{k}}^{\left(n\right)} \frac{p_{v c \mathbf{k}}^\alpha}{E_{cv\mathbf{k}}}d^2 \mathbf{k}, \label{eq:defs_0}
\end{equation}
and
\begin{equation}
	Q_{n m}^\alpha = i \int \,\psi_{cv\mathbf{k}}^{\left(n\right)*}\left[\psi_{cv\mathbf{k}}^{\left(m\right)}\right]_{; k_\alpha}d^2 \mathbf{k}, \label{eq:defs}
\end{equation}
where $ \left[\psi_{cv\mathbf{k}}^{\left(m\right)}\right]_{; k_\alpha} $ is the generalized derivative\cite{aversa_nonlinear_1995} in the $ \alpha $--direction of the exciton wave function for the state $ m $ given in terms of the Berry connection $ \Omega_{ij\mathbf{k}}^{\alpha} $, defined as\cite{pedersen_intraband_2015}
\begin{align}
	\left[\psi_{cv\mathbf{k}}^{\left(m\right)}\right]_{; k_\alpha}&=\frac{\partial \psi_{cv\mathbf{k}}^{\left(m\right)}}{\partial k_\alpha}-i\left(\Omega_{cc\mathbf{k}}^{\alpha}-\Omega_{vv\mathbf{k}}^{\alpha}\right)\psi_{cv\mathbf{k}}^{\left(m\right)}.\label{eq:generalized_deriv-exciton}
\end{align}
Analogously to what was discussed regarding Eq. (\ref{eq:generalized_deriv-momentum}), the excitonic wave function will be independent of $k_z$ and, as such, the $\frac{\partial}{\partial k_z}\psi_{cv\mathbf{k}}^{\left(m\right)}$ term is dropped, meaning that $Q_{n m}^z$ reads
\begin{align}
	Q_{n m}^z= \int \,\psi_{cv\mathbf{k}}^{\left(n\right)*}\left(\Omega_{cc\mathbf{k}}^{z}-\Omega_{vv\mathbf{k}}^{z}\right)\psi_{cv\mathbf{k}}^{\left(m\right)}d^2 \mathbf{k}.
\end{align}
Additionally, one can easily convert Eqs. (\ref{eq:excitonic_first-order_simp1}-\ref{eq:excitonic_second-order_simp1}) into formulas for the associated susceptibility as $\chi_{\alpha\beta}=\frac{i}{\omega\epsilon_0}\sigma_{\alpha\beta}$ and $\chi_{\alpha\beta\gamma}^{\mathrm{SHG}}=\frac{i}{2\omega\epsilon_0 }\sigma_{\alpha\beta\gamma}^{\mathrm{SHG}}$, respectively\cite{pedersen_intraband_2015}.

\section{Optical Response of Buckled Gapped Graphene}\label{Sec:buckling}

We will now quickly outline the IP optical selection rules for gapped graphene with trigonal warping, already discussed in the literature\cite{taghizadeh_nonlinear_2019}, before focusing our attention on the OOP linear and non--linear optical response of buckled gapped graphene. 

As discussed in Sec. \ref{Sec:gapped}, we will be ignoring $ \pi - \sigma $ hybridization by assuming the buckling is much smaller than the lattice constant. Therefore, this model will be identical to the unbuckled monolayer discussed previously, apart from the alternating $ z $--positions of the individual sublattices. More importantly, the eigenstates will remain those given by Eqs. (\ref{eq:eigenvectors_fixed1}-\ref{eq:eigenvectors_fixed2}), meaning that no changes to either the momentum matrix elements or to the Bethe--Salpeter equation are needed.

Throughout, we will consider a buckling parameter $ h=a/4 $, where $ a $ is the lattice parameter. This matches approximately what is present in the literature\cite{le_fracture_2021,ge_comparative_2016,xu_first-principle_2017,chen_site_2016} where, depending on the material in question, the buckling parameter $ h $ takes values between $ a/2.5 $ and $ a/8.6 $. Additionally, as will be evident, the presence of trigonal warping is not necessary to obtain finite linear and non--linear OOP conductivities.

\subsection{In--Plane Optical Selection Rules}

To obtain the optical selection rules, we must compute the angular integrals present in the excitonic matrix elements $ X_{0 n}^x $ and $ Q_{n m}^x $. These optical selection rules are relevant not only for the IP linear and non--linear response, but also for the non--diagonal OOP response. For clarity, we separate this subsection into discussion on linear and non--linear response.
For the linear optical response, we focus on the angular integral present in the definition of $ X_{0 n}^x $ following Eq. (\ref{eq:defs_0}). 
	To zeroth order in $ \zeta_{\mathrm{TW}} $, the angular integral in Eq. (\ref{eq:defs_0}) then reads
	\begin{align}
		\int_{0}^{2\pi} \,e^{i \ell_n \theta}p_{v c \mathbf{k}}^x d\theta
		&\propto \left[\frac{\Delta}{\varepsilon}+\frac{\ell_n+\tau}{\left|\ell_n+\tau\right|}\tau
		\right]\delta_{\left|\ell_n+\tau\right|,1}.\label{eq:angular_X}
	\end{align}
	The presence of the Kronecker delta in Eq. (\ref{eq:angular_X}) immediately gives rise to the well--known valley--dependent selection rules in gapped graphene, hexagonal Boron Nitride, and other monolayer materials with an hexagonal lattice \cite{taghizadeh_nonlinear_2019,henriques_optical_2020,quintela_theoretical_2022}when one takes into account the valley--dependent pseudo-spin contribution.  Including trigonal warping effects would lead to a quadratic correction allowing for transitions with $ \left|\ell_n+\tau\right|=2 $ or $ \left|\ell_n+\tau\right|=4 $.
	
	For the non--linear optical response, we first focus our attention on the angular integral in the definition of $ Q_{n m}^x $ following Eq. (\ref{eq:defs}). 
	Performing the necessary angular integral, we write the matrix elements in a somewhat abusive but concise form as 
	\begin{align}
		Q_{n m}^x&=Q_{\left|\ell_{m,n}\right|=1}^x+\zeta_{\mathrm{TW}}\left[Q_{\left|\ell_{m,n}\right|=2}^x+Q_{\left|\ell_{m,n}\right|=4}^x\right],\label{eq:simp_Q}
	\end{align}
	where the new indices restrict each term to the Kronecker deltas resulting from the different angular integrals and we defined $ \ell_{m,n}=\ell_m -\ell_n $ for conciseness.
	Besides the $ Q_{n m}^x $ matrix element, a linear contribution in $ \zeta_{\mathrm{TW}} $ is also present in the expansion of $ X_{0 n}^x $.
	In the same notation, we write this contribution as 
	\begin{align}
		X_{0 n}^x &=X_{\left|\ell_n+\tau\right|=1}^x+\zeta_{\mathrm{TW}}\left[X_{\left|\ell_n+\tau\right|=2}^x+X_{\left|\ell_n+\tau\right|=4}^x\right]\label{eq:simp_X}.
	\end{align}
		
		As we are considering contributions only up to first order in $ \zeta_{\mathrm{TW}} $, we must carefully analyze the matrix product $ X_{0 n}^x Q_{n m}^x X_{m 0}^x $ to understand which states are to be included. Knowing the simplified forms for the matrix elements, we can expand the oscillator strength $ X_{0 n}^x Q_{n m}^x X_{m 0}^x $ up to linear order in $ \zeta_{\mathrm{TW}} $. The non--zero contributions to the non--linear second order conductivity then read
		\begin{align}
			&\zeta_{\mathrm{TW}}\left[X_{\left|\ell_n+\tau\right|=2}^xQ_{\left|\ell_{m,n}\right|=1}^xX^{x,*}_{\left|\ell_m+\tau\right|=1}+\right.\nonumber\\
			&\qquad+X_{\left|\ell_n+\tau\right|=1}^xQ_{\left|\ell_{m,n}\right|=2}^xX^{x,*}_{\left|\ell_m+\tau\right|=1}+\nonumber\\
			&\qquad\left.+X_{\left|\ell_n+\tau\right|=1}^xQ_{\left|\ell_{m,n}\right|=1}^xX^{x,*}_{\left|\ell_m+\tau\right|=2}\right],\label{eq:shg_separated}
		\end{align}
			where the importance of including trigonal warping in order to obtain a non--zero second--order response is evident. Defining the oscillator strength $ \sigma_{\ell_n;\ell_m}\equiv X_{\ell_n}^x Q_{\ell_n,\ell_m}^x X_{\ell_m}^x $ from the allowed transitions of Eq. (\ref{eq:shg_separated}), the dominant matrix elements correspond to the $ \sigma_{p_+;s}$ and $\sigma_{s;p_+}$, in perfect agreement with Fig. (2-c) of Ref. \cite{taghizadeh_nonlinear_2019}.

\begin{figure*}
	\begin{minipage}[c]{0.7\textwidth}
		\hspace{-0.75cm}\includegraphics[scale=0.93403]{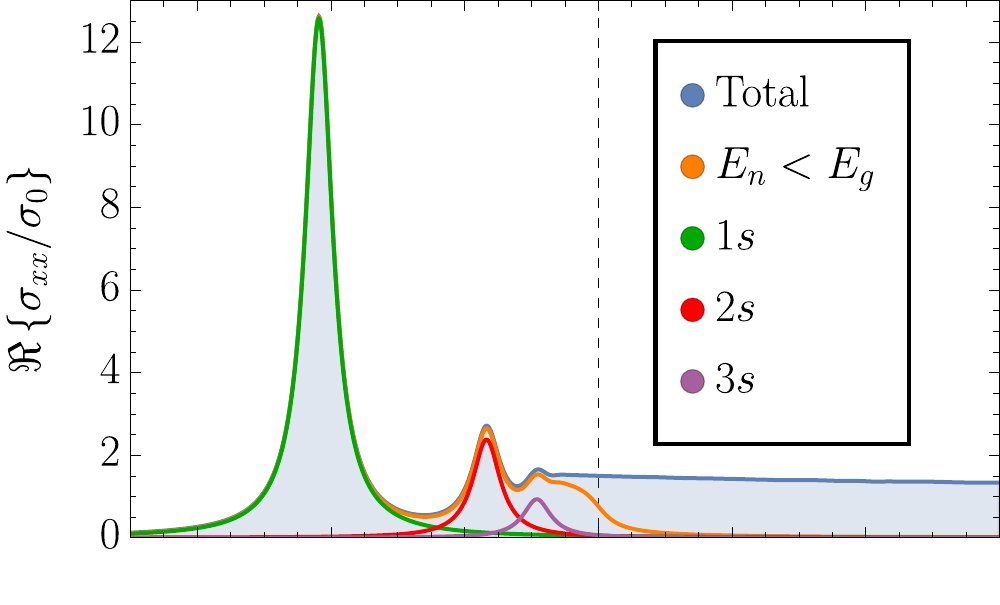}
		
		\vspace{-0.67cm}\hspace{-0.645cm}\includegraphics[scale=0.98]{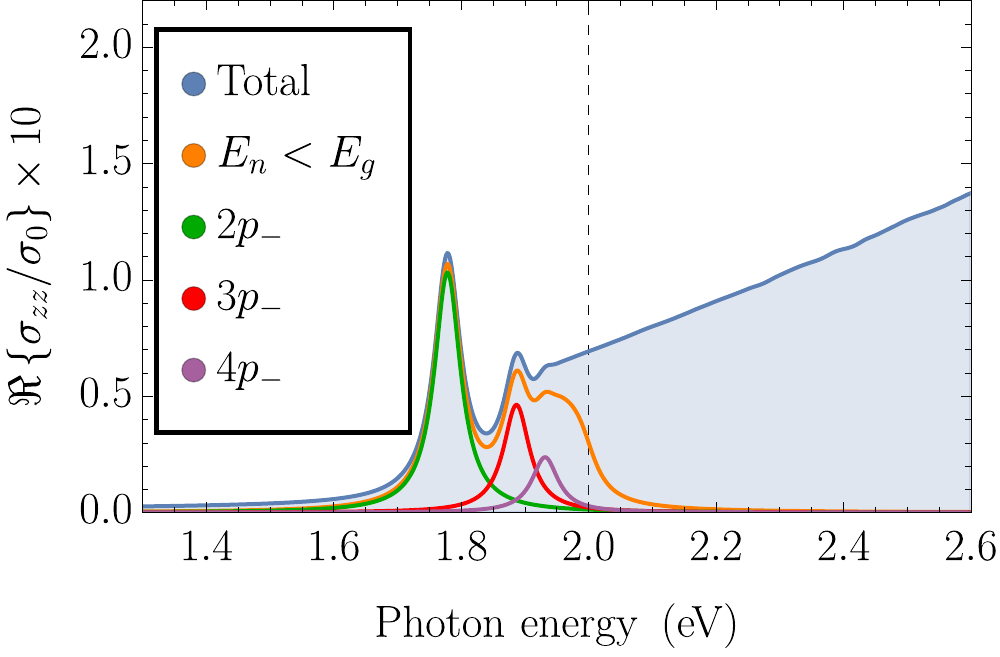}
	\end{minipage}%
	\begin{minipage}[c]{0.3\textwidth}
		\vspace{-1.5cm}\includegraphics[scale=0.7]{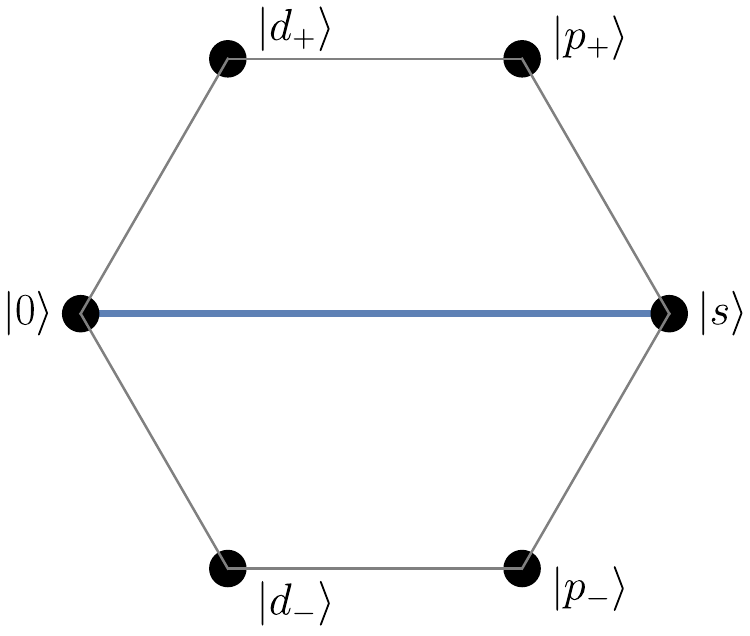}	
		
		\vspace{0.8cm}\includegraphics[scale=0.7]{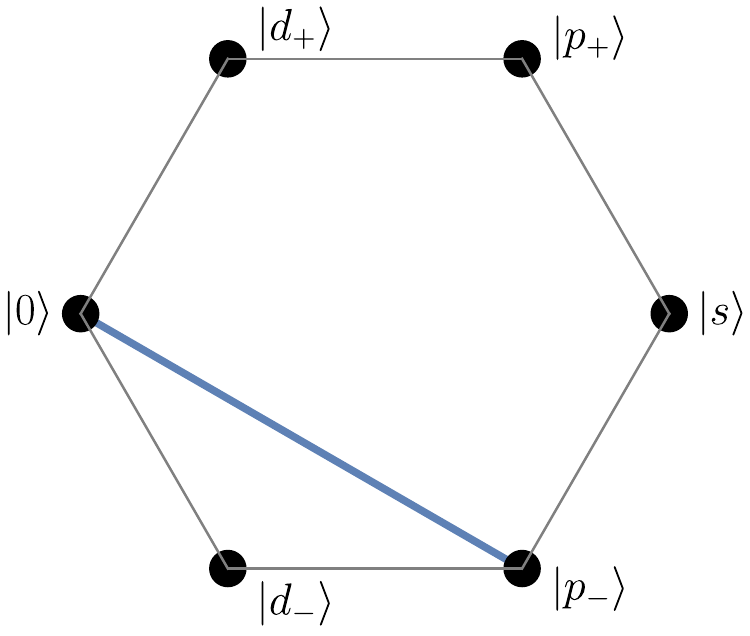}
	\end{minipage}
	\centering
	\caption{(Left) Real part of the linear IP (top) and OOP (bottom) optical response for $ \epsilon=1 $. 
		Orange curve corresponds to the excitonic bound states, while blue line also includes continuum states. Vertical dashed line represents the bandgap. 
		(Right) Diagram of dominant excitonic selection rules in the $\tau=1$ valley for linear IP (top) and OOP (bottom) optical response. 
		\label{fig:linear-excitonic-both}}
\end{figure*}			

\subsection{Out--of--Plane Momentum and Berry Connection}

The matrix elements of $ z $ are given by $ h\sigma_z $, with $ \sigma_z $ the diagonal Pauli matrix, and can be easily computed between bands $ n $ and $ m $ as 
\begin{align}
	z_{nm\mathbf{k}}&=\left\langle n_\mathbf{k}\middle|\left[
	\begin{array}{cc}
		h & 0\\
		0 & -h
	\end{array}
	\right]\middle|m_\mathbf{k}\right\rangle.\label{eq:z_vc_not-approx}
\end{align}
Under the same linear approximation in $ \zeta_{\mathrm{TW}} $ for the band structure as discussed in Eq. (\ref{eq:bandstructure-fixed}) and considering only terms up to $ \mathcal{O}\left(\zeta_{\mathrm{TW}}^1\right) $, Eq. (\ref{eq:z_vc_not-approx}) reads
\begin{align}
	z_{vc\mathbf{k}}&=-h e^{i\tau\theta}\sqrt{1-\frac{\Delta^2}{\varepsilon^2}}\left[1+\zeta_{\mathrm{TW}}\tau \frac{\Delta^2}{\varepsilon^2}a k \sin 3 \theta\right],\label{eq:z_vc_approx}\\
	z_{cc\mathbf{k}}&=h\frac{\Delta}{\varepsilon}\left[1-\zeta_{\mathrm{TW}}  \tau \left(1-\frac{\Delta^2}{\varepsilon^2}\right)a k\sin 3\theta\right]\nonumber\\
	&=-z_{vv\mathbf{k}}\label{eq:z_vv_approx}
\end{align}
for the different band pairs.

Knowing the $ z_{ij\mathbf{k}} $ matrix elements, we can finally write the OOP component of the momentum and Berry connections as 
\begin{align}
	&p_{vc\mathbf{k}}^{z}=\frac{m}{i\hbar}2\varepsilon z_{vc\mathbf{k}}\nonumber\\
	&\;\;=2ih\frac{m}{\hbar}e^{i\tau\theta}\sqrt{\varepsilon^2-\Delta^2}\left[1+\zeta_{\mathrm{TW}}\tau \frac{\Delta^2}{\varepsilon^2}a k \sin 3 \theta\right]\label{eq:p_vc_out-of-plane}
\end{align}
and, following from \cite{aversa_nonlinear_1995},
\begin{align}
	&\Omega_{cc\mathbf{k}}^{z}-\Omega_{vv\mathbf{k}}^{z}=z_{cc\mathbf{k}}-z_{vv\mathbf{k}}\nonumber\\
	&\quad=2h\frac{\Delta}{\varepsilon}\left[1-\zeta_{\mathrm{TW}} a k \tau  \left(1-\frac{\Delta^2}{\varepsilon^2}\right)\sin 3\theta\right].\label{eq:Omega_vc_out-of-plane}
\end{align}
The jump from $\frac{\partial}{\partial k_z}$ to $i z$ can be understood by considering the buckled monolayer as a repeated structure in the $ z$--direction. This means that the wavefunctions carry a $e^{i k_z z}$ factor, while the periodic parts (\emph{i.e.}, the eigenvectors in Eqs. (\ref{eq:eigenvectors_fixed1}-\ref{eq:eigenvectors_fixed2})) are independent of $k_z$. Finally, the period of this repeated structure is taken to infinity.

While the IP momentum in Eq. (\ref{eq:angular_X}) goes to zero as $ \Delta/\varepsilon\approx k^{-1}$ for large $k$, the dominant term of $ p_{vc\mathbf{k}}^{z} $ is linear in $ k $. As a consequence, contributions from continuum states (\emph{i.e.}, states where $ E_{n} > 2\Delta $) will quickly increase with $ \omega $. 

\subsection{Out--of--Plane Excitonic Linear Conductivity}

We will now analyze the OOP excitonic conductivity. Considering only the zeroth order contribution from $ \zeta_{\mathrm{TW}} $, it is immediately evident from the OOP momentum of Eq. (\ref{eq:p_vc_out-of-plane}) that only transitions to excitonic states with $ \ell_n=-\tau $ are allowed, meaning that $ X_{0 n}^z $ reads
\begin{align}
	X_{0 n}^z &=- \frac{2\pi hm}{\hbar}\delta_{\ell_n,-\tau}\int_{0}^{\infty} \,f_{cvk}^{(n)}\sqrt{1-\frac{\Delta^2}{\varepsilon^2}}kdk.\label{eq:simp_X_OOP}
\end{align}
Including trigonal warping effects would allow for transitions where $ \left|\ell_n+\tau\right|=3 $ and the correction would be quadratic in $ \zeta_{\mathrm{TW}}$.

The real part of the linear excitonic optical conductivity of the buckled monolayer for $ \epsilon=1 $ is plotted in Fig. (\ref{fig:linear-excitonic-both}), with the top panel the IP response and the bottom panel the OOP response. Right side diagrams represent the transitions allowed for each component in the $\tau=1$ valley. Considering the $\tau=-1$ valley would imply a sign flip of the diagrams (\emph{e.g.}, exchanging $p_-$ by $p_+$, etc), as evident from the selection rules of Eqs. (\ref{eq:simp_X}) and (\ref{eq:simp_X_OOP}). As expected from the form of the momentum operator $ p_{vc}^{z} $, we observe an ever--increasing linear optical conductivity when accounting for continuum states (see Appendix \ref{app:electronic_conductivities}). The optical response present in the bottom panel of Fig. (\ref{fig:linear-excitonic-both}) also qualitatively matches the measured optical conductivity of anisotropic materials, such as $\mathrm{ZrSiS}$, $\mathrm{ZrGeS}$, and $\mathrm{ZrGeSe}$, found in the current literature\cite{PhysRevLett.127.076402,PhysRevB.106.075143}.

\subsection{Out--of--Plane Excitonic Non--Linear Conductivity}

\begin{figure*}
	\begin{minipage}[c]{0.7\textwidth}
		\hspace{-0.5cm}\includegraphics[scale=1]{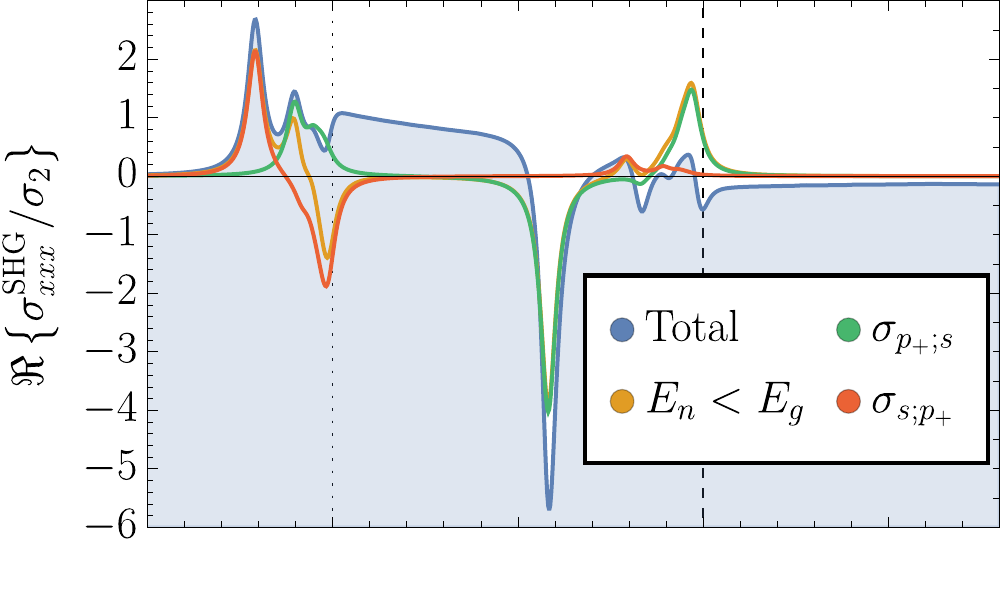}
		
		\vspace{-0.71cm}\hspace{-0.5cm}\includegraphics[scale=1]{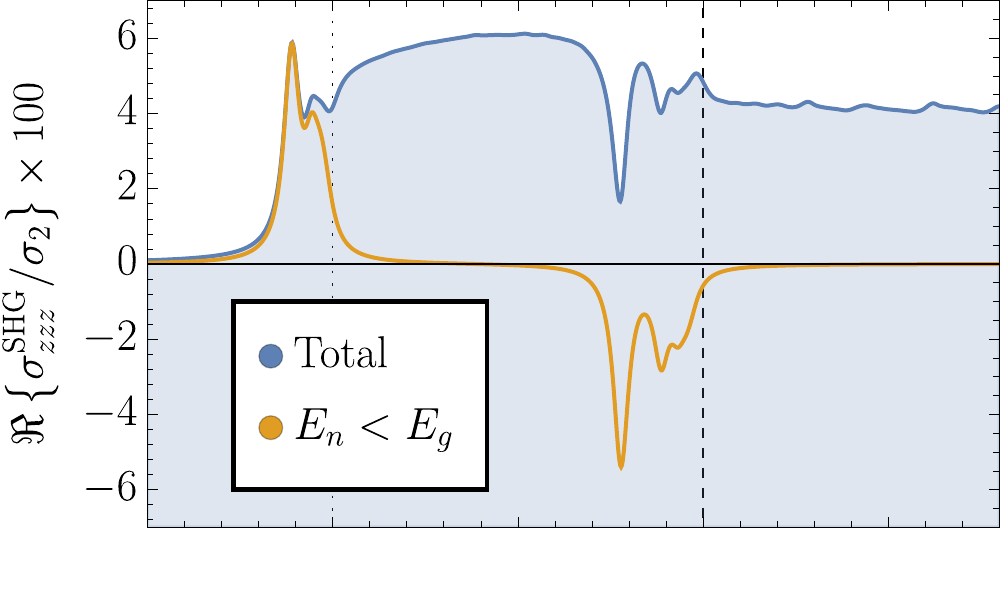}	
		
		\vspace{-1.01cm}\hspace{-0.5cm}\includegraphics[scale=1]{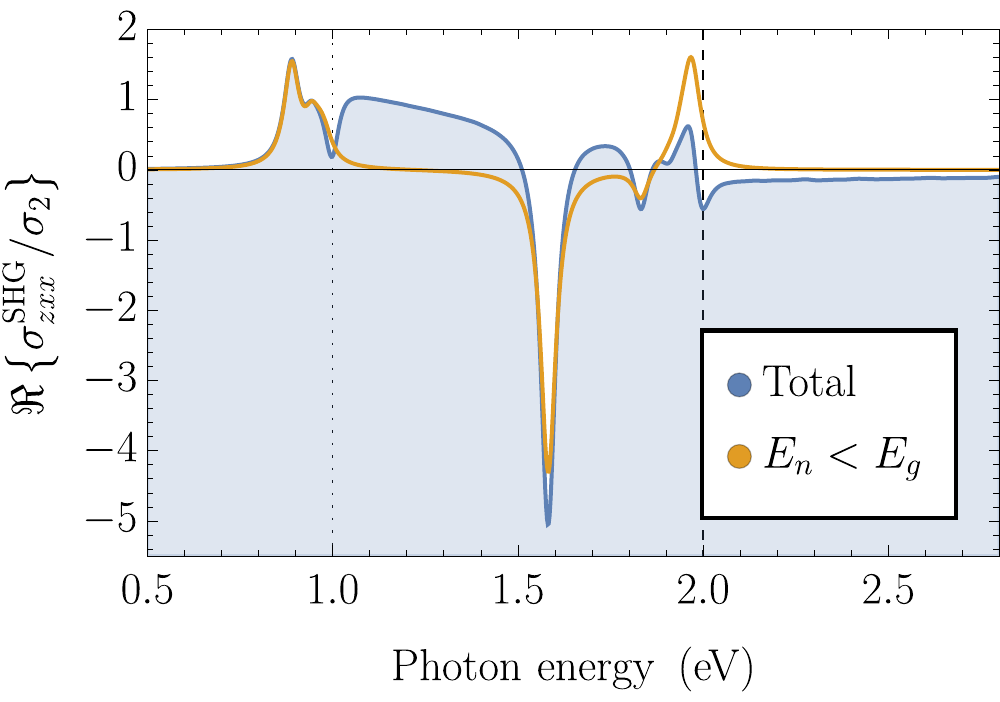}
	\end{minipage}%
	\begin{minipage}[c]{0.3\textwidth}
		\vspace{-1.5cm}\includegraphics[scale=0.7]{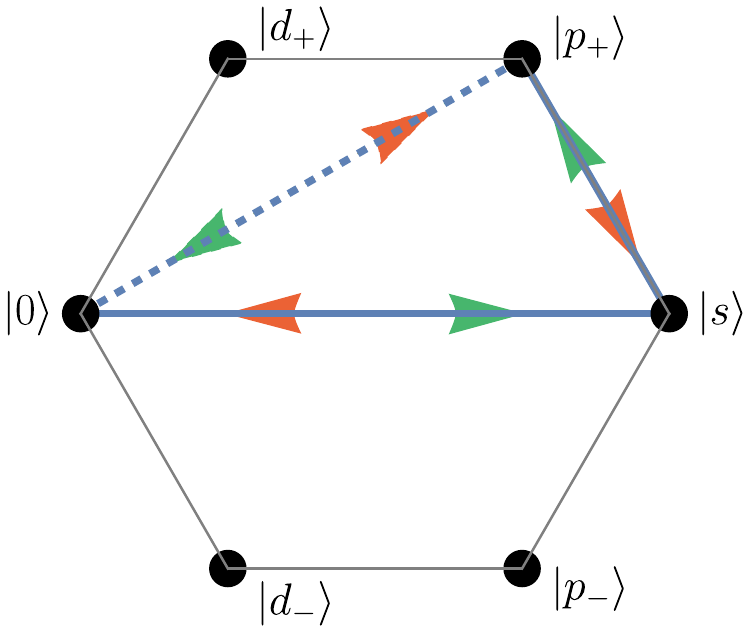}
		
		\vspace{0.8cm}\includegraphics[scale=0.7]{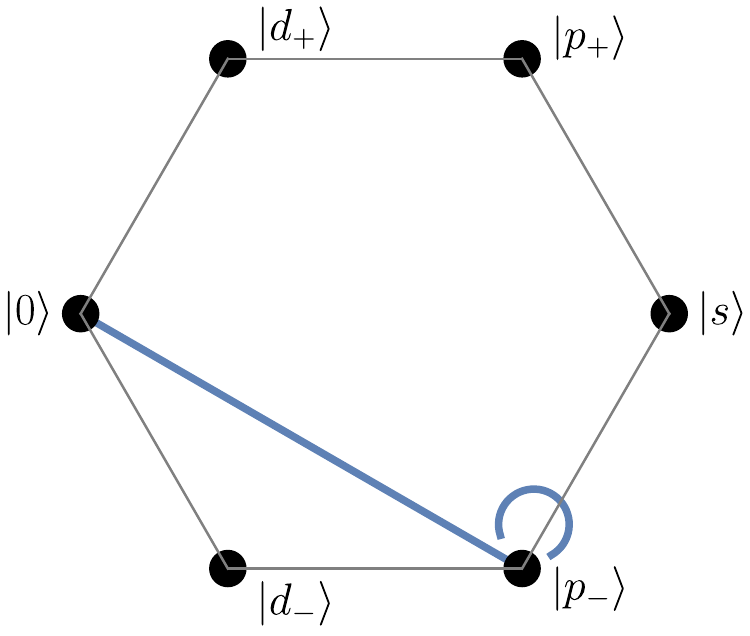}
		
		\vspace{0.8cm}\includegraphics[scale=0.7]{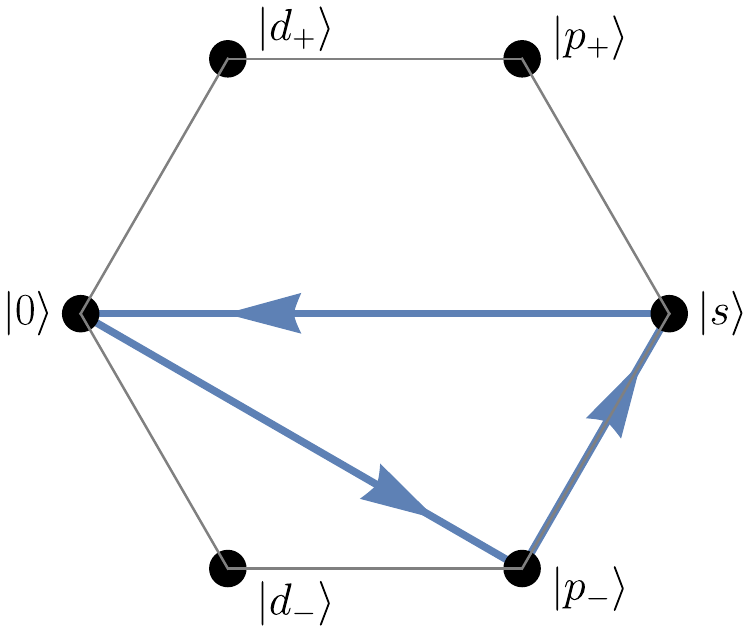}
	\end{minipage}
	\centering
	\caption{(Left) Real part of the SHG optical response with diagonal IP (top), diagonal OOP (middle) and non--diagonal OOP (bottom) conductivity for $ \epsilon=1 $, $ h=a/4 $. Orange curve corresponds to only excitonic bound states, while blue line also includes continuum states. Vertical (dotted) dashed black lines represent (half) the bandgap of the system. 
		(Right) Diagram of dominant excitonic selection rules in the $\tau=1$ valley for each component. Dashed line means the  transition is allowed by trigonal warping, solid lines are transitions allowed without trigonal warping. Arrow direction and colour represent the specific resonance when multiple contributions are present. 
		\label{fig:nonlinear-excitonic-both}}
\end{figure*}

Focusing now on the non--linear regime and considering only zeroth order in $ \zeta_{\mathrm{TW}} $ contributions, the $ Q_{n m}^z $ matrix element reads
\begin{align}
	Q_{n m}^z 
	&=4\pi h\delta_{\ell_m,\ell_n}\int_0^\infty \,f_{cvk}^{(n)*}\frac{\Delta}{\varepsilon}f_{cvk}^{(m)}kdk,\label{eq:simp_Q_OOP}
\end{align}
allowing transitions between states with the same angular momentum. However, as  $ X_{0 n}^z $ only allows $ \ell_n=-\tau $ to zeroth order in $ \zeta_{\mathrm{TW}} $, we arrive at the fact that only $ \ell_n=\ell_m=-1 $ states contribute in the $ \tau=1 $ valley. 

Including trigonal warping effects in $ Q_{n m}^z $ would allow for transitions where $ \left|\ell_m-\ell_n\right|=3 $. Considering this extra term together with the selection rules present in $ X_{0 n}^z $ leads to a vanishing first order contribution from $ \zeta_{\mathrm{TW}} $ to the SHG conductivity. Additionally, as each momentum matrix element will carry a factor of $ h $, the SHG conductivity will therefore be proportional to $ \left(h/a\right)^3 $. The SHG conductivity is plotted in the middle panel of Fig. (\ref{fig:nonlinear-excitonic-both}) for $ \epsilon=1 $. Apart from the much smaller amplitude due to the cubic dependence on $h/a$, it is also noteworthy that the response above $\hbar \omega=2\,\mathrm{eV}$ remains remarkably close to its maximum value. This is very different from what occurs for the IP response, where the response above $\hbar \omega=2\,\mathrm{eV}$ is much smaller than its maximum value. 

\subsection{Non--Diagonal Out--of--Plane Response}

Finally, we will consider the non--diagonal OOP response in buckled gapped graphene. Considering again only the $ x $--direction for the IP response, we have three different components which can prove interesting: $ \sigma_{zxx} $, $ \sigma_{xzx} $ and $ \sigma_{zzx} $

Looking more carefully at the selection rules of the system, we can immediately tell that $ \sigma_{zzx}=0 $ when recalling Eqs. (\ref{eq:simp_X},\ref{eq:simp_X_OOP},\ref{eq:simp_Q_OOP}): while $ {X}_{0 n}^z $ and $ {Q}_{n m}^z $ only allow $ \ell_n=\ell_m=-1 $ states, $ X_{0 n}^x $ explicitly forbids these states. As such, we focus our attention only on $ \sigma_{zxx} $. Although we will not be discussing $ \sigma_{xzx} $, 
a quick analysis of the various selection rules discussed previously shows that the dominant response shall be a zeroth order contribution in $ \zeta_{\mathrm{TW}} $ of the form $X_{\ell_n=0}^x Q_{nm}^z X_{\ell_m=0}^x$. Under Kleinman symmetry\cite{PhysRev.128.1761}, $ \sigma_{xzx} $ will be approximately equal to $ \sigma_{zxx} $.

\begin{figure}
	\hspace{-0.2cm}\includegraphics[scale=0.7]{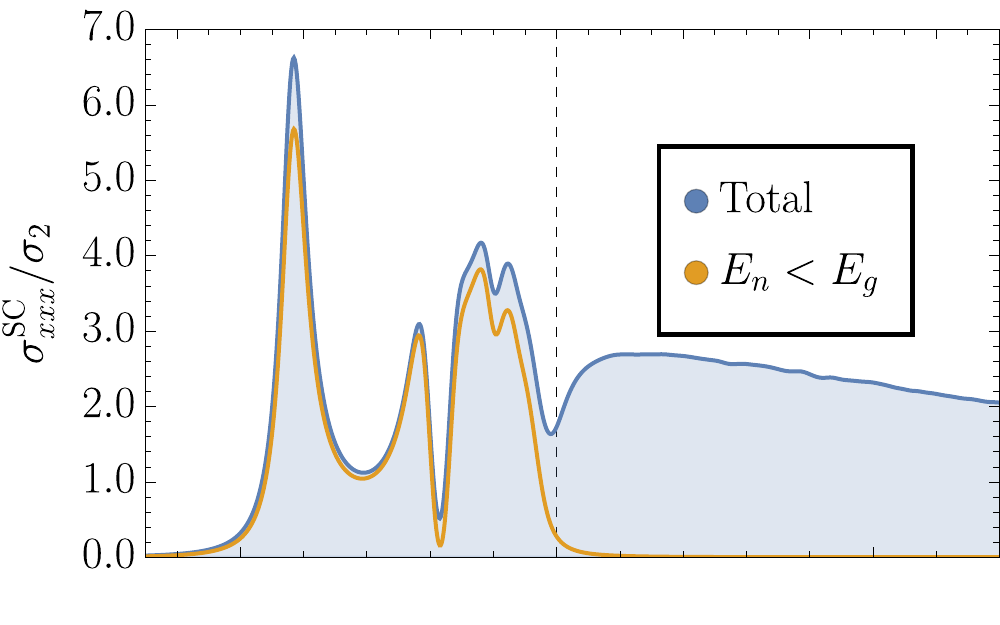}
	
	\vspace{-0.519cm}\hspace{-0.2cm}\includegraphics[scale=0.7]{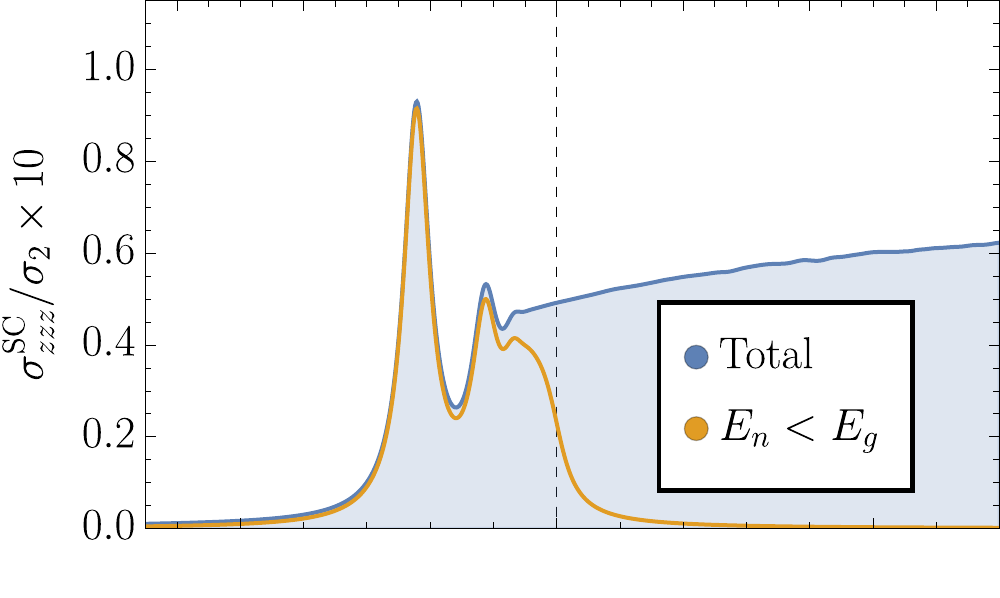}	
	
	\vspace{-0.527cm}\hspace{-0.2cm}\includegraphics[scale=0.7]{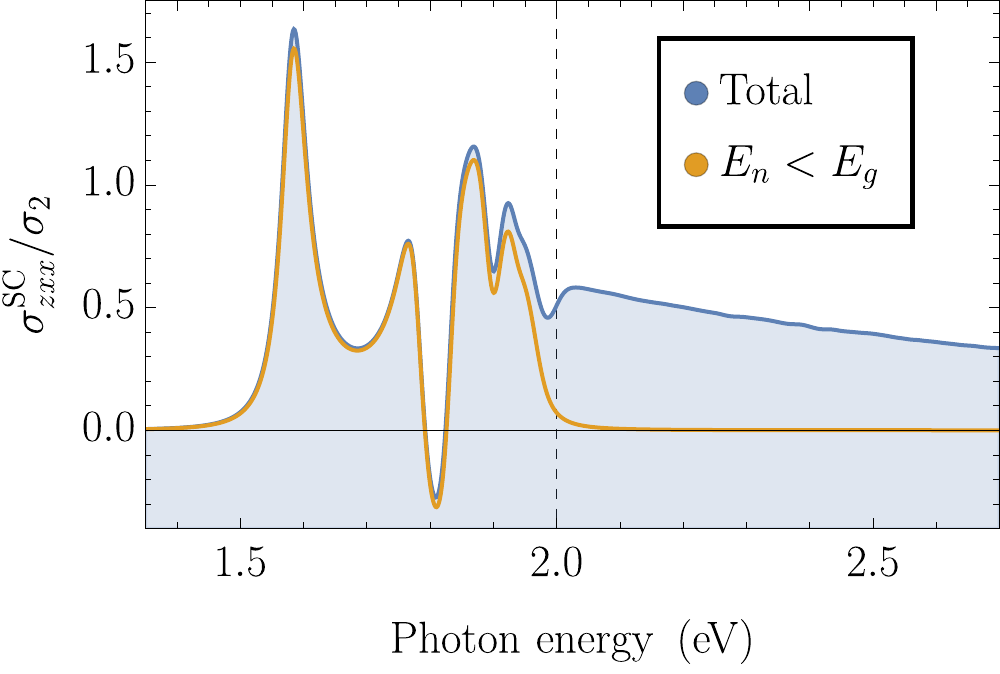}
	
	\caption{Non--linear SC IP (top), diagonal OOP (middle) and non--diagonal OOP (bottom) optical for $ \epsilon=1 $, $ h=a/4 $. Orange curve corresponds to only excitonic bound states, while blue line also includes continuum states. Vertical dashed black lines represents the bandgap of the system.
		\label{fig:SC_excitonic_panel}}
\end{figure}

Now explicitly computing the selection rules for $ \sigma_{zxx} $, $ X_{0 n}^z $ again immediately forces $ \ell_n=-1 $. Recalling Eqs. (\ref{eq:simp_Q},\ref{eq:simp_X}), 
the dominant transition will be associated with the matrix elements $Q_{\left|\ell_{m,n}\right|=1}^x X_{\left|\ell_m+\tau\right|=1}^x$ (\emph{i.e.}, zeroth order contribution in $ \zeta_{\mathrm{TW}} $), meaning that $ \ell_m $ is restricted to $ \left|\ell_m+\tau\right|=1 $.  
Excluding all other contributions, we can immediately expect that this off--diagonal term will be significantly larger than $ \sigma_{zzz}^{(2)} $, as the dependence on $ h/a <1 $ will be linear instead of cubic. Additionally, $ X_{m 0}^x $ is much larger than $ X_{m 0}^z $, which will also contribute to this trend. 

This excitonic non--linear conductivity is then plotted in the bottom panel of Fig. (\ref{fig:nonlinear-excitonic-both}) for $ \epsilon=1 $. As expected from the qualitative analysis of the matrix elements, the relative magnitude of the off--diagonal OOP contribution is much larger than the diagonal OOP response present in the top panel of Fig. (\ref{fig:nonlinear-excitonic-both}). As discussed previously, this mainly stems from the lower order dependence in $ h/a $.
Additionally, and as expected from the general form $ \sigma_{z xx} $, we can also observe that the bound state peaks corresponding to $ 2\hbar\omega = E_n $ (\emph{i.e.}, states below $ \hbar\omega=\Delta $) match exactly with the corresponding regime in $ \sigma_{z z z} $, while those corresponding to $ \hbar\omega = E_m $ match exactly with the same regime in $ \sigma_{x x x} $.

We also observe that the magnitude of $ \sigma_{zxx}^{\mathrm{S H G}} $ is remarkably close to that of $ \sigma_{xxx}^{\mathrm{S H G}} $ for the buckling parameter chosen. This will, of course, be dictated by the ratio $h/a$, meaning that for a larger buckling parameter the non--diagonal OOP SHG response will be larger than the diagonal IP SHG response. Additionally, in Fig. (\ref{fig:SC_excitonic_panel}), we plot the SC for the three different tensor elements discussed previously. As the selection rules are the same as presented in Fig. (\ref{fig:nonlinear-excitonic-both}), they are not included in the panels. 

\begin{figure}
	\hspace{-0.5cm}\includegraphics[scale=0.75]{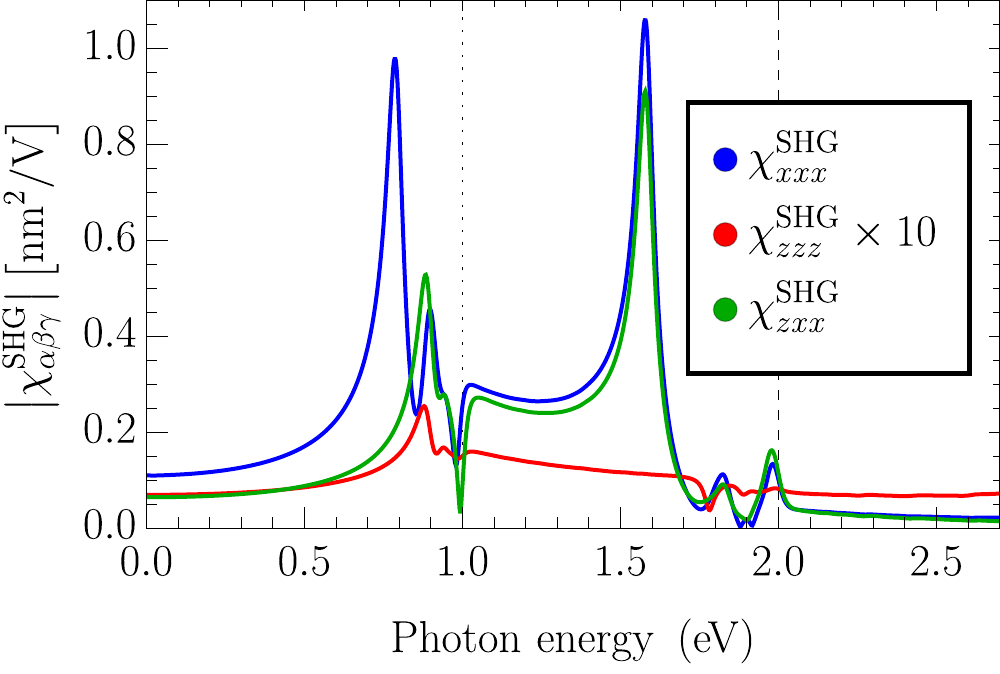}
	\caption{Magnitude of three different components ($x x x$, $zzz$ and $zxx$) of SHG non--linear optical susceptibilities for $\epsilon=1$, $h=a/4$. Vertical (dotted) dashed black lines represent (half) the bandgap of the system. 
		\label{fig:susceptibility}}
\end{figure}

Finally, we present the $x x x$, $zzz$ and $zxx$ components of the absolute value of the SHG non--linear optical susceptibilities. These can be directly computed from the conductivity as 
\begin{align}
	\chi_{\alpha\beta\gamma}^{\mathrm{SHG}}&=\frac{i}{2\omega\epsilon_0 }\sigma_{\alpha\beta\gamma}^{\mathrm{SHG}}\label{eq:susc}
\end{align}
and their absolute value is presented in Fig. (\ref{fig:susceptibility}). Due to the inclusion of a finite broadening $\hbar\Gamma=0.05\,\mathrm{eV}$, the three considered tensor elements of the conductivity take a small but non--zero value at $\hbar\omega=0$, with its magnitude less than one percent of the maximum of each tensor component. Still, the presence of this finite value at $\hbar\omega=0$ means that the broadening must also be considered in the $1/\omega$ factor present in Eq. (\ref{eq:susc}).

The relative amplitudes of the different components can be easily compared, with $\chi_{zxx}$ presenting a very similar amplitude to $\chi_{xxx}$. Additionally, $\chi_{zzz}$ is roughly a factor of $1/20$ smaller than either $\chi_{xxx}$ or $\chi_{zxx}$ within the bandgap of the system, as expected from the cubic dependence on the ratio $h/a$. The different dependence on $h/a$ in each component means that, as discussed previously, choosing a larger buckling parameter will lead to a comparatively greater OOP susceptibility. Notably, the left--most peak of $\chi_{xxx}$ is not present in $\chi_{zxx}$. This is due to the different selection rules for the two components of the SHG non--linear susceptibility, where certain transitions present in $\chi_{xxx}$ are no longer allowed for $\chi_{zxx}$.

\section{Summary}

In this paper, we studied the excitonic linear and non--linear optical properties of anisotropic buckled monolayer semiconductors. To this end, we began by considering the gapped Dirac model with trigonal warping. The excitonic states were computed by numerical diagonalization of the Bethe--Salpeter equation, allowing us to explicitly discuss the excitonic selection rules of the system.

Introducing a small buckling in the lattice structure of the monolayer, we then obtained the OOP momentum matrix elements and Berry connections, discussing the resulting OOP excitonic optical selection rules. We then analyzed the $x x x$, $zzz$ and $zxx$ tensor elements of both SHG and SC optical response, discussing the differences and similarities between the three components. 

Finally, we computed the absolute value of the non--linear optical susceptibility, directly comparing the amplitudes of the $\chi_{xxx}$, $\chi_{zzz}$ and $\chi_{zxx}$ matrix elements. The OOP magnitudes are, of course, dictated by the ratio between the buckling parameter $(h)$ and the lattice constant $(a)$, meaning that a  structure with a different buckling parameter will present greatly different relative magnitudes. While the OOP diagonal component had a much smaller maximum amplitude, stemming from the cubic dependence on the ratio $h/a$, the non--diagonal OOP component had a very similar amplitude to that of the diagonal IP component. 

\section*{Acknowledgements}

M.F.C.M.Q. acknowledges the International Iberian Nanotechnology Laboratory (INL) and the Portuguese Foundation for Science and Technology (FCT) for the Quantum Portugal Initiative (QPI) grant SFRH/BD/151114/2021. 

\begin{widetext}
\appendix
\section{Electronic Linear and Non--Linear Conductivity Expressions \label{app:electronic_conductivities}}

In this appendix, we will present the expressions for the free--carrier conductivity in our monolayer system. These were computed directly from the definitions in Eqs. (\ref{eq:generic_electronic}-\ref{eq:generic_electronic-nl}) while considering only contributions up to first order in $ \zeta_{\mathrm{TW}} $. For this effect, we recall the definitions of the in-- and OOP momentum matrix elements and Berry connections from Ref. \cite{taghizadeh_nonlinear_2019} (with the appropriate gauge transformation) and Eqs. (\ref{eq:p_vc_out-of-plane},\ref{eq:Omega_vc_out-of-plane}). 

In the following expressions $E_{cv\mathbf{k}}=E_{cvk}$, as including the contribution from trigonal warping in the bandstructure would introduce contributions one order higher in $\zeta_{\mathrm{TW}}$ which would then vanish upon integration and summation over valley. As such, the band structure now only depends on the radial component $k$, meaning that $E_{cvk}=2\varepsilon$, where $\varepsilon$ is as defined in Eq. (\ref{eq:short_var}). As an example, the non--linear IP response would, up to first order, include an extra term originating from $p_{v c\mathbf{k}}^\alpha\left[p_{c v\mathbf{k}}^\beta\right]_{; k_\lambda}$ expanded to zeroth order, which vanishes upon integration.

\begin{figure}
	\centering
	\hspace{-0.75cm}\includegraphics[scale=0.72]{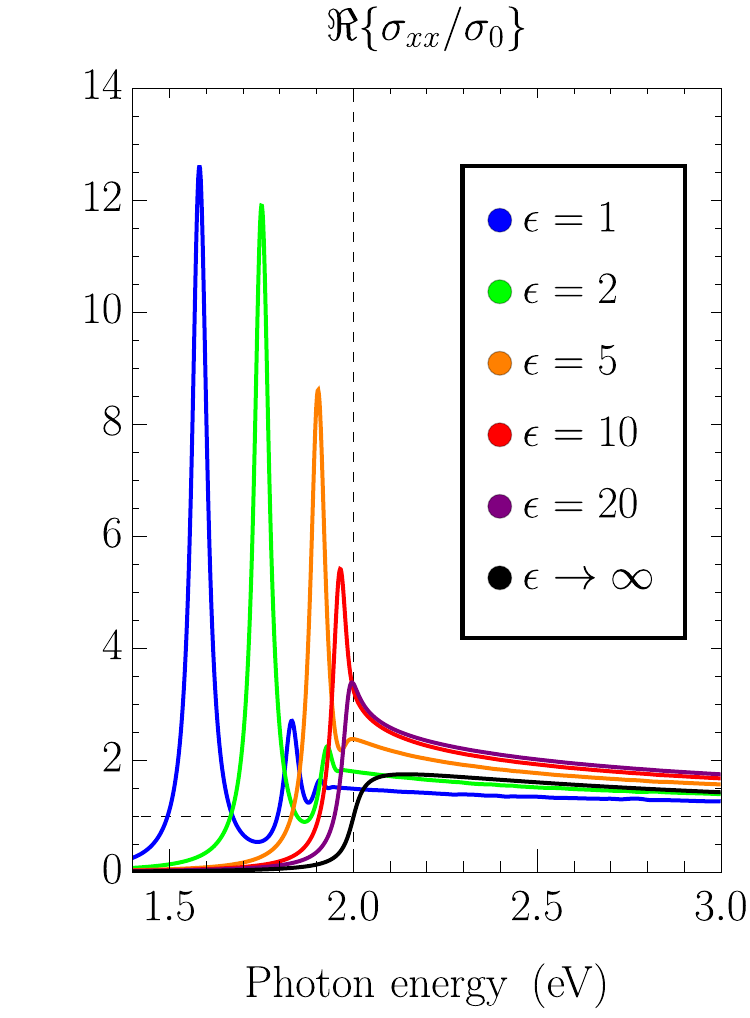}\hspace{-0.5cm}\includegraphics[scale=0.7]{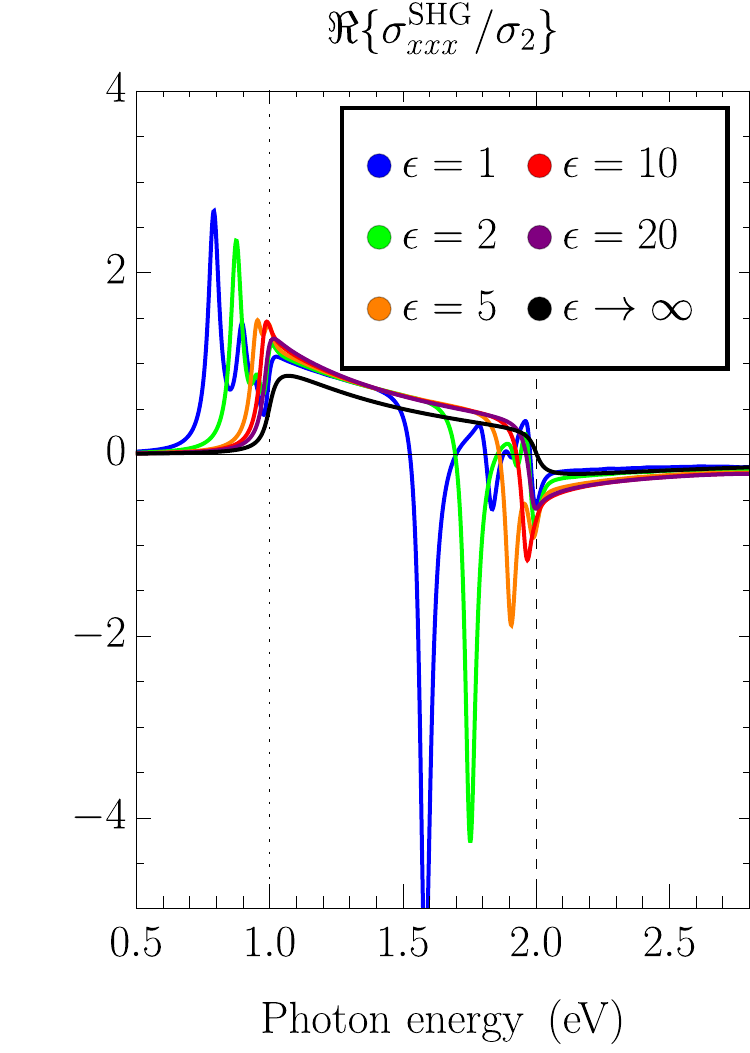}
	\hspace{-0.2cm}\includegraphics[scale=0.67]{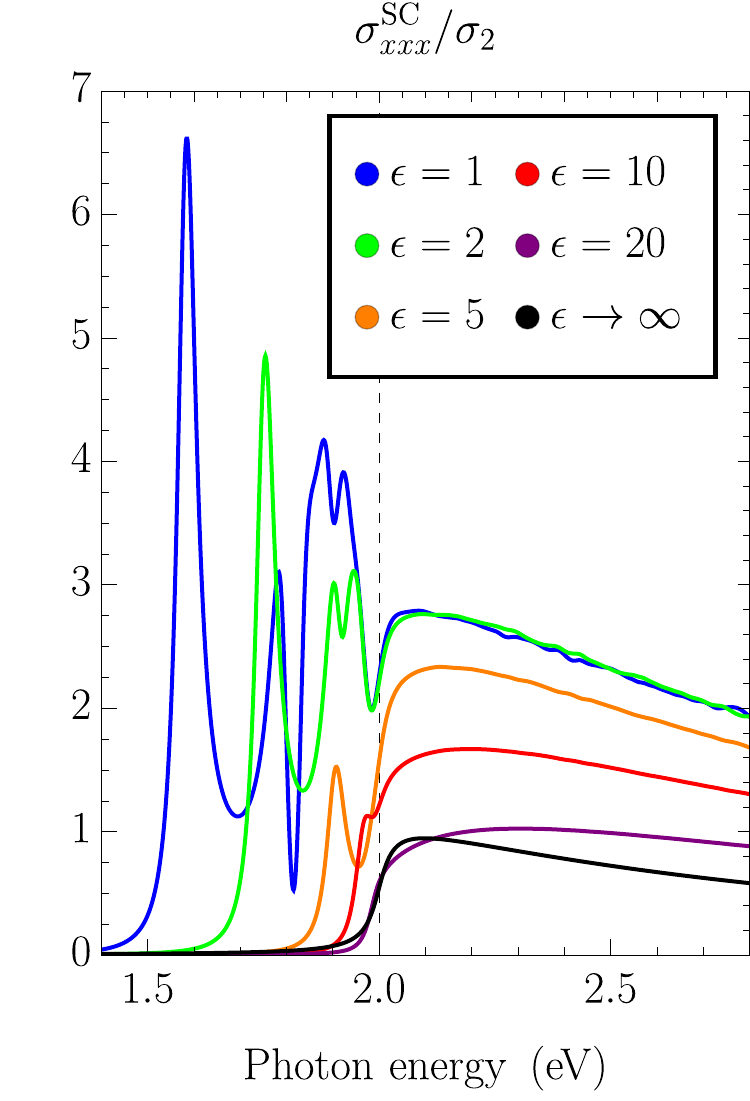}
	\caption{Convergence of the real part of the (left) linear, (middle) SHG and (right) SC IP optical response towards the free carrier limit as the dielectric constant $ \epsilon $ increases. Vertical axis is in units of $ \sigma_{0} $ for the linear response and $ \sigma_{2} $ for the non--linear response. \label{fig:convergence1}}
\end{figure}

Starting with the diagonal linear response described in Eq. (\ref{eq:generic_electronic}), it follows that 
\begin{align}
	\frac{\sigma_{x x}(\omega)}{\sigma_0}&=\frac{2 i }{\pi}\left[\frac{E_g}{\hbar \omega}-\left(1+\frac{E_g^2}{\hbar^2 \omega^2}\right) \operatorname{tanh}^{-1}\left(\frac{\hbar \omega}{E_g}\right)\right],
\end{align}
where $ E_g = 2\Delta $ and $ \operatorname{tanh}^{-1} $ is the inverse hyperbolic tangent.

Under a similar analysis, we compute the angular integral present in Eq. (\ref{eq:generic_electronic-nl}) and obtain the generic radial integral form of the diagonal second order response as 
\begin{align}
	\frac{\sigma_{x x x}^{(\mathrm{i n t r a})}\left(\omega_p, \omega_q\right)}{\sigma_2}&=i  \frac{4 \zeta_{\mathrm{T W}}}{\pi} E_g^2 \int_{E_g}^{\infty} \frac{\hbar \omega_p+\hbar \omega_{p q}}{\left(E_{cvk}^2-\hbar^2 \omega_q^2\right)\left(E_{cvk}^2-\hbar^2 \omega_{p q}^2\right)} d E_{cvk}+(p \leftrightarrow q) .
\end{align}
Choosing specifically SHG
 and SC
 processes, we obtain
\begin{align}
	\frac{\sigma_{x x x}^{\mathrm{SHG}}(\omega)}{\sigma_2} &\equiv \frac{\sigma_{x x x}^{(\mathrm{i n t r a})}(\omega, \omega)}{\sigma_2}=i  \frac{8 \zeta_{\mathrm{T W}}}{\pi}\left(\frac{E_g}{\hbar \omega}\right)^2\left[\operatorname{tanh}^{-1}\left(\frac{\hbar \omega}{E_g}\right)-\frac{1}{2} \operatorname{tanh}^{-1}\left(\frac{2 \hbar \omega}{E_g}\right)\right]
\end{align}
and
\begin{align}
	\frac{\sigma_{x x x}^{\mathrm{SC}}(\omega)}{\sigma_2} &\equiv \frac{\sigma_{x x x}^{(\mathrm{intra })}\left(\omega,-\omega^*\right)}{\sigma_2}= \frac{16 \zeta_{\mathrm{T W}}}{\pi} \Im\left[\left(\frac{E_g}{\hbar \omega}\right)^2 \operatorname{tanh}^{-1}\left(\frac{\hbar \omega}{E_g}\right)-\frac{E_g}{\hbar \omega}\right],
\end{align}
where $\Im$ denotes the imaginary part.

A similar analysis can be done for the diagonal OOP linear and non--linear response, although one must compute $ \sigma_{zz} $ carefully as the integration leads to a divergent result if the infinite $ k $--space is considered. This is, however, only true for the imaginary part. Restricting our analysis to the real part, we find a finite result reading 
\begin{align}
	\Re \left[\frac{\sigma_{z z}(\omega)}{\sigma_{0}}\right] \approx \frac{8}{3 \gamma^2}\left(\frac{h}{a}\right)^2\left(\hbar^2 \omega^2-E_g^2\right) H\left(\hbar \omega-E_g\right),
\end{align}
where $ H(x) $ represents the Heaviside step function. For the diagonal non--linear response, no convergence issues are present and the integral can be considered over the infinite $ k $--space 
\begin{align}
	\frac{\sigma_{z z z}^{(\mathrm{i n t r a})}\left(\omega_p, \omega_q\right)}{\sigma_2}= \frac{32 \hbar\left(\omega_p+\omega_{p q}\right) E_g^2}{3 i \pi \gamma^2}\left(\frac{h}{a}\right)^3 \int_{E_g}^{\infty} \frac{E_{cvk}^2-E_g^2}{\left(E_{cvk}^2-\hbar^2 \omega_p^2\right)\left(E_{cvk}^2-\hbar^2 \omega_{p q}^2\right)} d E_{cvk}+(p \leftrightarrow q).
\end{align}
Again restricting our analysis to SHG
 and SC,
 we obtain 
\begin{align}
	\frac{\sigma_{z z z}^{\mathrm{SHG}}(\omega)}{\sigma_2}=\frac{32 E_g^2}{3 i \pi \gamma^2}\left(\frac{h}{a}\right)^3\left[\left(\frac{E_g^2}{\hbar^2 \omega^2}-1\right) \operatorname{tanh}^{-1}\left(\frac{\hbar \omega}{E_g}\right)-\frac{1}{2}\left(\frac{E_g^2}{\hbar^2 \omega^2}-4\right) \operatorname{tanh}^{-1}\left(\frac{2 \hbar \omega}{E_g}\right)\right]
\end{align}
and
\begin{align}
	\frac{\sigma_{z z z}^{\mathrm{SC}}(\omega)}{\sigma_{2}}= \frac{32 E_g^2}{\pi \gamma^2}\left(\frac{h}{a}\right)^3 \Im\left[\frac{E_g}{\hbar \omega}+\left(1-\frac{E_g^2}{\hbar^2 \omega^2}\right) \operatorname{tanh}^{-1}\left(\frac{\hbar \omega}{E_g}\right)\right].
\end{align}

\begin{figure}
	\centering
	\hspace{-0.75cm}\includegraphics[scale=0.75]{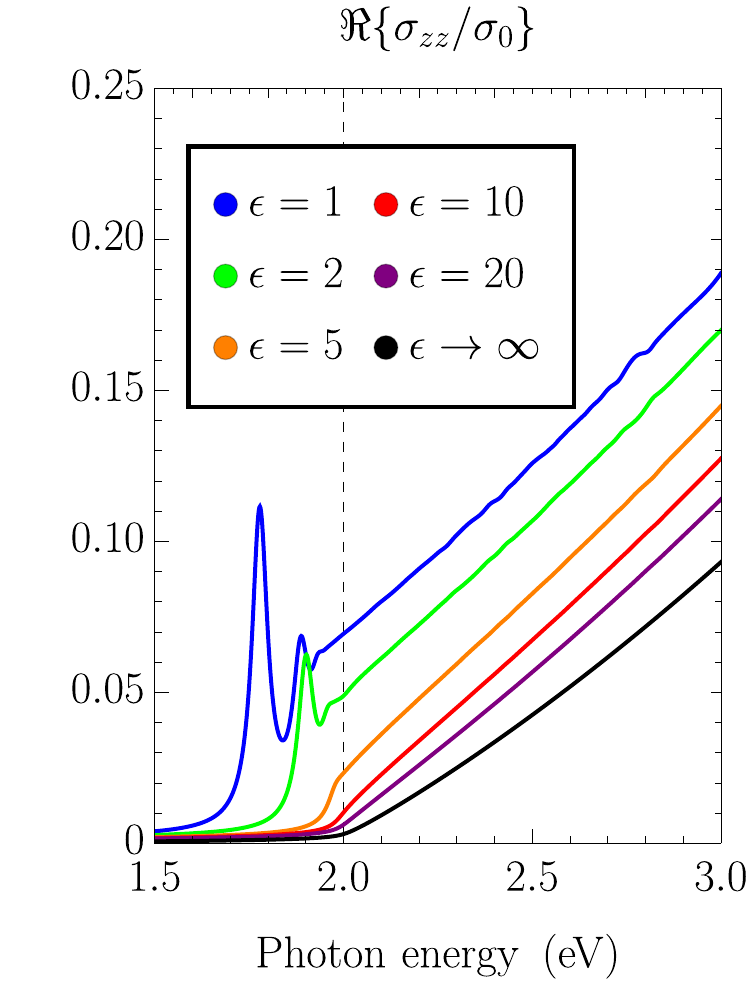}\hspace{-0.6cm}\includegraphics[scale=0.72]{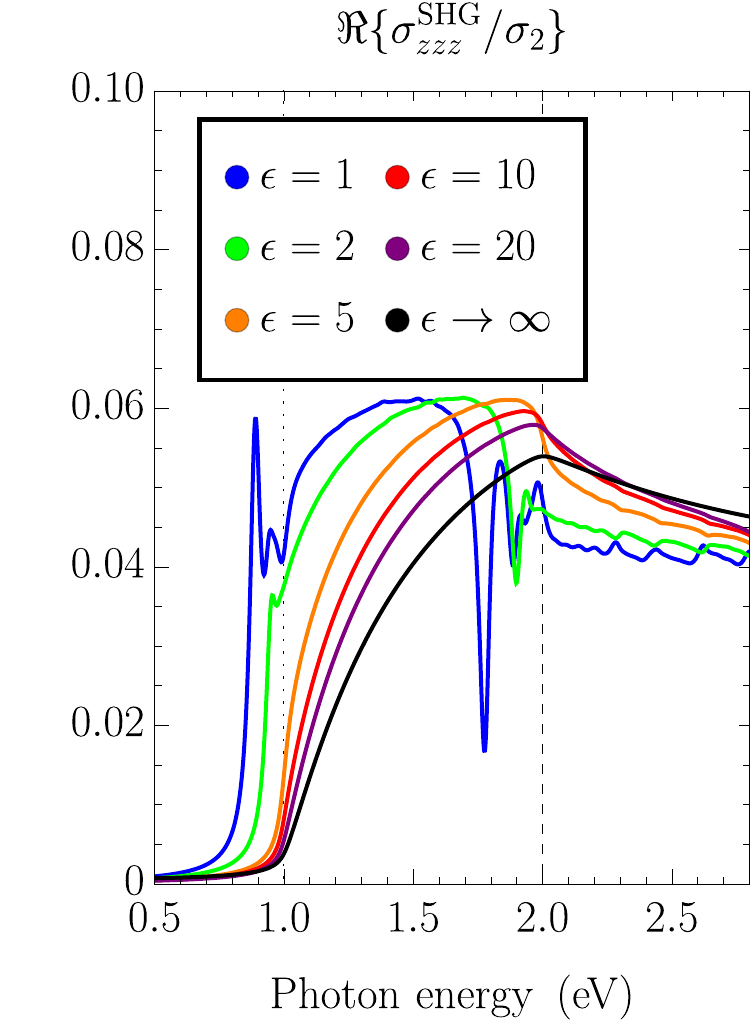}\hspace{-0.3cm}\includegraphics[scale=0.72]{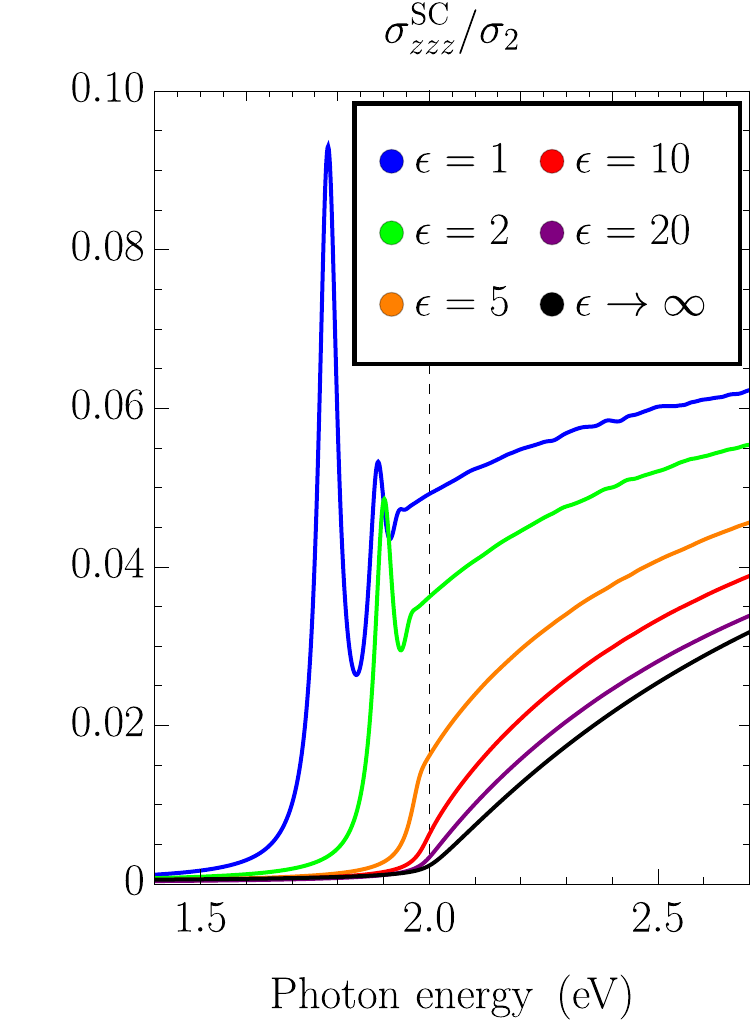}
	\caption{Same as Fig. (\ref{fig:convergence1}), except for the diagonal OOP response. \label{fig:convergence2}}
\end{figure}

Finally, we consider the off--diagonal OOP response $ \sigma_{zxx} $, where we can also consider the integral over infinite $ k $--space, reading
\begin{align}
	\frac{\sigma_{z xx}^{(\mathrm{i n t r a})}\left(\omega_p, \omega_q\right)}{\sigma_2}= 16 i\frac{ \left(\hbar \omega_p+\hbar \omega_{p q}\right)}{\pi}\Delta^2 \frac{h}{a} \int_{E_g}^{\infty}  \frac{E_g^2-E_{cvk}^2}{E_{cvk}^2\left(E_{cvk}^2-\hbar^2 \omega_q^2\right)\left(E_{cvk}^2-\hbar^2 \omega_{p q}^2\right)} d E_{cvk}+(p \leftrightarrow q).
\end{align}
Looking again at SHG and SC, we obtain
\begin{align}
	\frac{\sigma_{z xx}^{\mathrm{SHG}}(\omega)}{\sigma_2}&=i\frac{1}{\pi} \frac{h}{a}\frac{E_g^2}{\hbar^2 \omega^2} \left[3 \frac{E_g}{\hbar \omega}-2\left(\frac{E_g^2}{\hbar^2 \omega^2}-1\right) \operatorname{tanh}^{-1}\left(\frac{\hbar \omega}{E_g}\right)+\frac{1}{2}\left(\frac{E_g^2}{\hbar^2 \omega^2}-4\right) \operatorname{tanh}^{-1}\left(\frac{2\hbar \omega}{E_g}\right)\right]
\end{align}
and
\begin{align}
	\frac{\sigma_{z xx}^{\mathrm{SC}}(\omega)}{\sigma_{2}}&= \frac{4}{\pi}\frac{h}{a} \Im\left[\frac{E_g^3}{\hbar^3 \omega^3}-\frac{2 E_g}{3 \hbar \omega}+\frac{E_g^2}{\hbar^2 \omega^2}\left(1-\frac{E_g^2}{\hbar^2 \omega^2}\right) \operatorname{tanh}^{-1}\left(\frac{\hbar \omega}{E_g}\right)\right].
\end{align}
Knowing these expressions, we can study the convergence of the excitonic conductivities towards the free--carrier regime as the dielectric constant increases. This is plotted in Figs. (\ref{fig:convergence1},\ref{fig:convergence2},\ref{fig:convergence3}) for dielectric constant $\epsilon$ between $ 1 $ and $ 20 $, as well as the free--carrier limit (in black). In these plots, we can see the excitonic conductivity converging towards the free--carrier regime as the dielectric constant of the medium surrounding the monolayer increases, as expected from the fast drop of binding energies and number of bound states.

\begin{figure}
	\centering
	\hspace{-0.5cm}\includegraphics[scale=0.7]{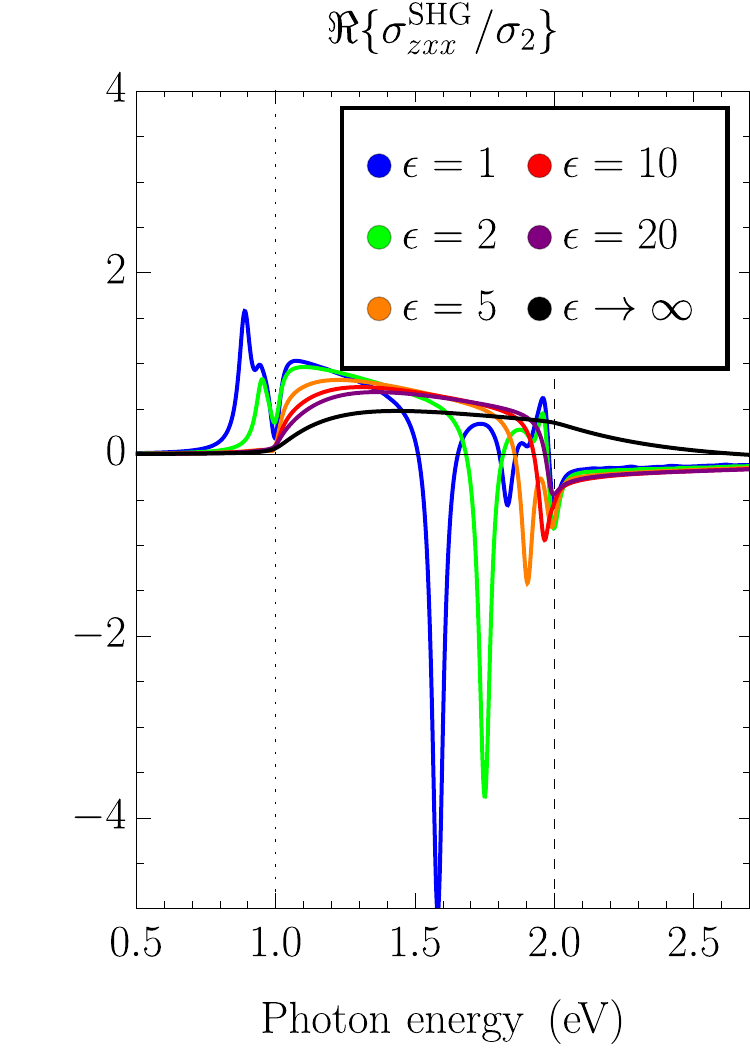}\hspace{0.5cm}\includegraphics[scale=0.7]{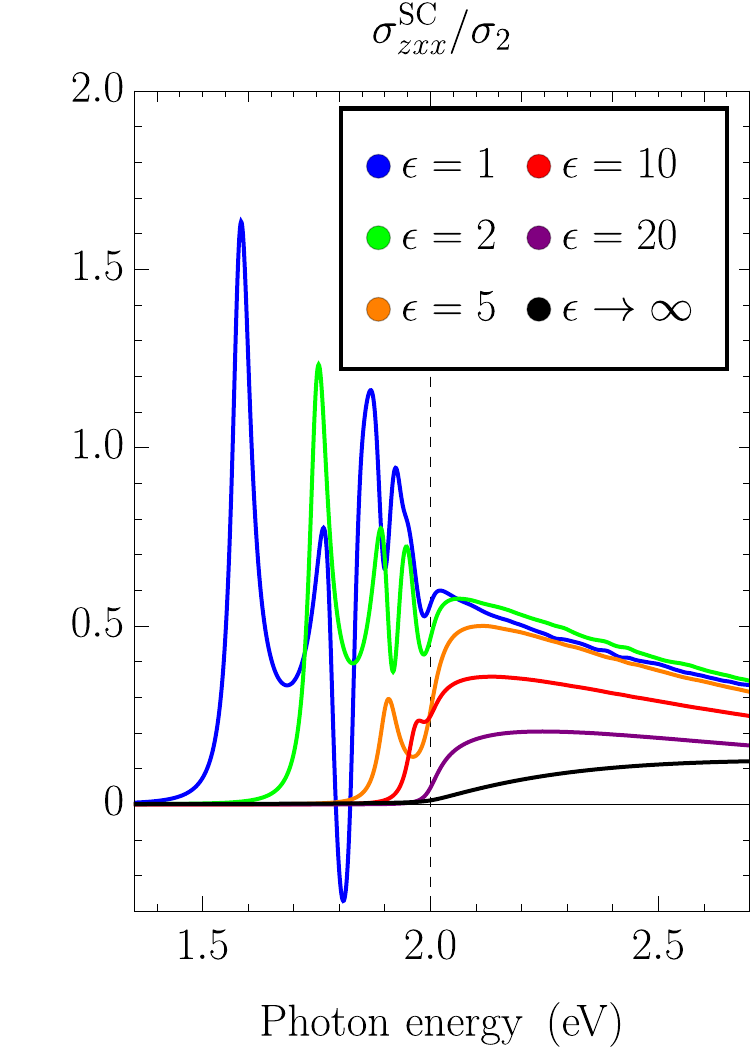}
	\caption{Convergence of the real part of the off--diagonal OOP SHG (left) and SC (right) $ zxx $ optical response towards the free carrier limit as the dielectric constant $ \epsilon $ increases. Vertical axis is in units of $ \sigma_{2} $. \label{fig:convergence3}}
\end{figure}
\end{widetext}

\bibliographystyle{ieeetr}
\bibliography{gapped-graphene_biblio}
\end{document}